\documentclass[a4paper,11pt]{article}
\usepackage{jcappub}
\usepackage{lineno}
\usepackage{rotating}

\hypersetup{linkcolor=magenta,citecolor=blue,filecolor=cyan,urlcolor=cyan}

\usepackage{amsmath, cancel, xcolor, hyperref}

\newcommand\penn{Department of Physics and Astronomy, University of Pennsylvania, Philadelphia, PA 19104, USA}
\newcommand\uiuc{Department of Astronomy, University of Illinois Urbana-Champaign, 1002 West Green Street, Urbana, IL, 61801, USA}
\newcommand\caltech{Department of Astronomy, California Institute of Technology, Pasadena, CA 91125, USA}
\newcommand\caps{Center for AstroPhysical Surveys, National Center for Supercomputing Applications, University of Illinois, Urbana, IL 61801, USA}
\newcommand\chicago{Department of Astronomy and Astrophysics, University of Chicago, Chicago, IL, 60637, USA}
\newcommand\arizona{Steward Observatory and Department of Astronomy, University of Arizona, Tucson, AZ 85721, USA}
\newcommand\uva{Department of Astronomy, University of Virginia, Charlottesville, VA 22904, USA}

\newcommand\ciins{[CII]}
\newcommand\cii{\ciins\;}
\newcommand\ciiumns{\ciins-158$\mu$m}
\newcommand\ciium{\ciiumns\;}
\newcommand\kperp{{k_{\perp}}}
\newcommand\kpara{{k_{\parallel}}}
\newcommand\kperpmin{{k_{\perp, \text{min}}}}
\newcommand\kperpmax{{k_{\perp, \text{max}}}}
\newcommand\kparamin{{k_{\parallel, \text{min}}}}
\newcommand\kparamax{{k_{\parallel, \text{max}}}}
\newcommand\halphans{H$\alpha$}
\newcommand\halpha{\halphans\;}
\newcommand\sonens{\texttt{A1}}
\newcommand\sfourns{\texttt{A4}}
\newcommand\shundns{\texttt{A100}}
\newcommand\sone{\sonens\;}
\newcommand\sfour{\sfourns\;}
\newcommand\shund{\shundns\;}

\graphicspath{{./}{figures/}{./figs}}

\arxivnumber{2602.XXXXX}
\title{
A New Window into the Baryon Cycle at Cosmic Noon with Line Intensity Mapping: Forecasts for auto- and cross-correlations in \ciiumns, HI 21 cm, CO$_{J+1\rightarrow J}$, and H$\alpha$ galaxies
}

\author[a, 1]{Shubh Agrawal,\note{Corresponding author}}
\author[a]{James E. Aguirre,}
\author[b,c,a]{Justin S. Bracks,}
\author[d]{Ryan P. Keenan,}
\author[c,e]{Charles M. Bradford,}
\author[f]{Brockton S. Brendal,}
\author[g]{Peter Dow,}
\author[f,h,i]{Jeffrey P. Filippini,}
\author[j,h]{Jianyang Fu,}
\author[i]{Karolina Garcia,}
\author[c,e]{Reinier M.J. Janssen,}
\author[g]{Bradley R. Johnson,}
\author[k]{Wooseok Kang,}
\author[l]{Christos Karoumpis,}
\author[m]{Garrett K. Keating,}
\author[a]{Adam Lidz,}
\author[c]{Lun-Jun Liu,}
\author[n]{Ian Lowe,}
\author[a]{Alexander Manduca,}
\author[o]{Aashrita Mangu,}
\author[n]{Daniel P. Marrone,}
\author[n]{Evan C. Mayer,}
\author[f]{Sydnee O'Donnell,}
\author[p,e]{Talia Saeid,}
\author[h]{Mathilde Van Cuyck,}
\author[f,h,i]{Joaquin Vieira,}
\author[k,q,r]{and Jessica A. Zebrowski}

\affiliation[a]{\penn}
\affiliation[b]{Department of Physics and Astronomy, University of California, Los Angeles, CA, 90095, USA}
\affiliation[c]{\caltech}
\affiliation[d]{Max-Planck-Institut für Astronomie, Königstuhl 17, D-69117 Heidelberg, Germany}
\affiliation[e]{Jet Propulsion Laboratory, California Institute of Technology, Pasadena, CA 91109, USA}
\affiliation[f]{Department of Physics, Grainger College of Engineering, University of Illinois Urbana-Champaign, Urbana, IL 61801, USA}
\affiliation[g]{\uva}
\affiliation[h]{\uiuc}
\affiliation[i]{\caps}
\affiliation[j]{Department of Physics, Princeton University, Princeton, NJ 08544, USA}
\affiliation[k]{\chicago}
\affiliation[l]{Argelander Institut für Astronomie, Universität Bonn, Auf dem Hügel 71, D-53121 Bonn, Germany}
\affiliation[m]{Center for Astrophysics $\vert$ Harvard \& Smithsonian, Cambridge, MA 02138, USA}
\affiliation[n]{\arizona}
\affiliation[o]{Department of Physics, University of Chicago, Chicago, IL, 60637, USA}
\affiliation[p]{School of Earth and Space Exploration, Arizona State University, Tempe, AZ 85287, USA}
\affiliation[q]{Fermi National Accelerator Laboratory, MS209, P.O. Box 500, Batavia, IL, 60510, USA}
\affiliation[r]{Kavli Institute for Cosmological Physics, University of Chicago, Chicago, IL, 60637, USA}

\emailAdd{shubh@sas.upenn.edu}

\abstract{
Across the peak of cosmic star formation at $z\sim1-2$, inflow, processing, and feedback drive rapid changes in the spatial distribution and chemical composition of baryons in galaxies and surrounding reservoirs; this baryon cycle can be tomographically mapped by line intensity mapping (LIM) of atomic hydrogen, ionized carbon, and carbon monoxide.

We present a simulation-based forecasting framework for detecting auto- and cross-power spectra between spectroscopic surveys of four such tracers at $z\sim0.5-1.7$ mapping the same deep field - the Terahertz Intensity Mapper, Epoch of Reionization Spectrometer on FYST, MeerKAT, and \textit{Euclid}.
We forward-model three-dimensional distributions for these tracers from magnetohydrodynamic simulations, directly capturing the two-halo, one-halo, and shot statistics without relying on analytical decompositions. We further detail a signal-to-noise formalism, tailored to LIM surveys with highly anisotropic geometries and Fourier-space coverage.

We demonstrate that galaxy cross-correlations will be the dominant discovery channel for current-generation surveys.
These instruments will detect the auto-spectra for CO and HI 21 cm and the CO $\times$ 21 cm cross-spectrum at modest S/N $\sim 1-10$, while placing upper limits on the \ciium signals. 
\ciins, CO, and HI LIM will be $\sim3-30\times$ ($0.5-1.5$ dex) more sensitive to cross-correlation with the \textit{Euclid} surveys, however, than their respective auto-correlations, constraining all three models of line emission at high significance (S/N $\sim 10-40$) within this decade.

Finally, we formulate a staged instrumental trajectory with planned or reasonable improvements, including the as-proposed mid-frequency Square Kilometer Array.
We forecast advancing the per-$k$-mode sensitivities of each of the auto-, galaxy-line, and line-line power spectra by several orders of magnitude,
enabling new percent- and sub-percent level constraints on cosmology and the redshift evolution of star formation and the baryon cycle.
}


\begin{document}
\maketitle
\flushbottom

\section{Introduction \label{sec:intro} }

As structures grew hierarchically from primordial fluctuations, two competing classes of dynamics were at play, each evolving over time. Dark matter and baryons accrete under gravitation, fueling star formation. The ensuing stellar winds and active galactic nuclei (AGN) expel gas and slow star formation, quenching growth at lower redshifts due to a decrease in available cold gas \citep{1978MNRAS.183..341W, 1980lssu.book.....P, 1986ApJ...303...39D, 1991ApJ...379...52W,1998A&A...331L...1S, 2005Natur.435..629S,2010ApJ...717..323B, 2015ARA&A..53...51S, Walter_2020}.
Multi-wavelength observations from the past decade indicate that the cosmic rate of star formation rose, from the epoch of reionization to its peak at roughly 3.5 Gyr after inflation, and then decayed by over an order of magnitude to the present day \citep{Madau_2014, F_rster_Schreiber_2020}. 
This ``cosmic noon" at $z\sim 1-2$ represents the epoch during which galaxies and their surrounding baryon reservoirs underwent their most rapid evolution and chemical transformation \citep{P_roux_2020, Walter_2020}. 
Understanding the interplay between the growth of large-scale structure and the chemical evolution of galaxies within the cosmic web is therefore one of the next frontiers of precision astrophysics and cosmology \citep{kovetz2017lineintensitymapping2017status, Cui_2017, kovetz2019astrophysicscosmologylineintensitymapping, vieira2020terahertzintensitymappertim, 2023A&A...676A..62V, Chung_2024, wright2024baryoncyclemoderncosmological, tojeiro2025large, dosibhatla2025tracinglargescalestructure}.

The redshift evolution of the baryon cycle through its different phases $-$ cold atomic hydrogen, molecular hydrogen, and hot ionized hydrogen $-$ has been measured by past galaxy surveys at multiple frequencies \citep{1998ARA&A..36..189K, 2012ARA&A..50..531K, Glazebrook_1999, 2008Ap&SS.313..253Y, 2013ARA&A..51..105C, 2013ARA&A..51..207B, delooze+14,  Madau_2014,2016ApJ...833...67W,Walter_2020, 2020ARA&A..58..363P, Maccagni_2024}. However, such surveys inherently miss the faint end of the luminosity function and diffuse baryon emission, as they only catalog discrete bright emitters that can be spatially resolved from their neighbors.
Line intensity mapping (LIM) is an emerging observational technique wherein narrow spectral features are tomographically measured to map the growth of large-scale structure and the evolution of galactic and extragalactic chemistry \citep{Fonseca_2016, kovetz2017lineintensitymapping2017status, kovetz2019astrophysicscosmologylineintensitymapping, Kewley_2019,fonseca2019lineintensitymappingnovel, Bernal_2022}. Measuring the aggregate emission evolution along the line of sight with a low-angular-resolution spectrometer, LIM surveys aim to provide a complete bulk census of a species, including faint and diffuse sources, with low sample variance. LIM is poised to be a next-generation probe of astrophysics and cosmology at epochs such as cosmic noon, where mapping multiple baryonic phases informs theories of galaxy formation. Simultaneously, percent and sub-percent measurements of the biased tracers of the matter power spectrum at these intermediate redshifts will provide complementary constraints on $\Lambda$CDM, characterizing dark matter, dark energy, and current cosmological tensions in measurements of the Hubble and S8 parameters \cite{kovetz2017lineintensitymapping2017status, kovetz2019astrophysicscosmologylineintensitymapping,  fonseca2019lineintensitymappingnovel, Bernal_2019, Bernal_2022, 2021CQGra..38o3001D, 2021ApJ...919...16F, 2023ARNPS..73..153K, 2024JCAP...12..007C, 2025JCAP...02..021A}.

While the emission of each line traces the underlying dark matter density field, at small scales in the one-halo and shot-noise-dominated regime, different lines encode information about aggregate emission and the distributions of distinct species. 
Four emission lines can together sample complementary phases and conversion processes within the baryon cycle; see Figure 1 from \cite{Schaan_2021} for an elucidating schematic.
The 21 cm spin-flip transition originates from neutral atomic hydrogen (HI) in the interstellar and intergalactic medium, providing information on diffuse gas reservoirs within the cosmic web \citep{furlanetto2019fundamentals21cmline, katz2024probingnewphysicscosmic}. 
The CO rotational ladder traces molecular gas (H$_2$) reservoirs in giant molecular clouds (GMCs), quantifying the cold fuel and potential for star formation \citep{1993prpl.conf..125B,1996PASJ...48..275A,2011A&A...528A.124C, 2012ApJ...746...69G, 2013ARA&A..51..207B, Narayanan_2014, Pe_aloza_2016}.
\ciiumns, an important coolant in the interstellar medium (ISM), is expected to be one of the brightest features from star-forming galaxies, arising from singly-ionized carbon (C$^+$) in predominantly photon-dominated regions (PDRs); it is therefore sensitive to the star-forming ISM and the evolution of galaxy star formation \citep{1985crawford,1991ApJStacey, 2021Yang, 2015Vallini, 2024Liang, Olsen_2017, delooze+14}. 
Similarly, \halpha emission originates largely in H\textsc{ii} regions, directly tracing the ionizing output of young stellar populations and hence active star formation rates (SFRs) \citep{1984ApJ...287..116K, Moustakas_2006, 2006agna.book.....O, Peimbert_2017, Schaan_2021, Tacchella_2022}.

Current LIM constraints on the intensities and power spectra of these emission lines around cosmic noon vary in precision and remain the focus of several ongoing and planned instruments \citep{battye2012bingosingledishapproach, 2014arXiv1412.4872D, 2016mks..confE..32S, 2016SPIE.9906E..5XN, 2017MNRAS.464.1948F, 2019ApJ...871...75I, 2019MNRAS.485.3486D, vieira2020terahertzintensitymappertim, Essinger_Hileman_2020, Karkare_2022, 2023ApJS..264....7C}. 
For the 21 cm line, bounds on its cross-correlation with galaxies and, more recently, its auto-correlation spectrum have been reported at $z \sim 1$ \citep{chang2010hydrogen21cmintensitymapping, 2013MNRAS.434L..46S, Masui_2013, Wolz_2021, paul2023detectionneutralhydrogenintensity, chimecollaboration2025detectioncosmological21cm}. The bulk and low-$J$ CO emission has been detected by the COPSS \citep{2015ApJ...814..140K} and mmIME \citep{2020ApJ...901..141K} at higher $z\gtrsim2.3$, indirectly constraining line intensity models near cosmic noon. Tentative detections of the mean \ciium intensity have also been reported $z\sim0.3-2.9$ from \textit{Herschel} \citep{agrawal2025farinfraredlineshiddenarchival} and at $z\sim 2.6$ from \textit{Planck} \citep{Pullen_2018, Yang_2019}, though they remain in tension.
Direct detection of \halpha line emission is widely used in galaxy counting surveys for extracting spectroscopic redshifts at $z\sim 1-2$, while SPHEREx and its potential successors (e.g., CDIM, \cite{cooray2016cosmicdawnintensitymapper}) aim to directly map \halpha at higher(lower) spectral(spatial) resolutions in their deep fields \citep{Gong_2017, Cheng_2024}. Current surveys have been limited by astrophysical and terrestrial foregrounds, interloper line contamination, high instrumental noise in relevant wavelength bands, and the fact that these lines comprise only a small fraction of the overall extragalactic background light in the infrared and radio.

With current experimental limitations, the merits of cross-correlating different tracers of the underlying dark matter density field have been emphasized repeatedly, particularly for mitigating foregrounds and systematics, including in the context of LIM \citep{Furlanetto_2007,Lidz_2008,Smith_2009,2013ApJ...768...15P,villaescusanavarro2015crosscorrelating21cmintensitymaps,Schaan_2021,Keenan_2022,Padmanabhan_2023,Fronenberg_2024}. Particularly compelling for cosmic noon studies is the boost in signal-to-noise obtained when cross-correlating with a tracer mapped in a wavelength regime that offers more favorable foregrounds, more mature and tractable detector technology, or existing robust instrumentation and engineering solutions. The benefits extend beyond observational advantages, however: these cross-correlations encode astrophysical information inaccessible to auto-correlations alone. Auto-power spectra cannot constrain the multivariate luminosity function and its conditional variants \citep{Yang_2003, Schaan_2021}, which inform models of joint halo occupation, galaxy-halo connection, and line-line luminosity covariance and correlated scatter. 
Scale- and redshift-dependent cross-correlation between lines quantifies the co-evolution of different baryonic phases, with implications for theories of star formation, galactic feedback, and chemical enrichment within halos. Additionally, the auto- and cross-power spectra encode different moments of the line luminosity functions in the shot regime (at small scales), while constraining distinct products of the mean line intensity and biases in the linear two-halo regime at large scales.

In this paper, we calculate the auto- and cross-correlation sensitivities of three classes of LIM surveys, corresponding to the current, next-generation, and a futuristic generation of instruments. 
The current-generation surveys plan to map the same deep field in the southern sky, creating substantial overlap in volume that enables robust cross-correlations, suppression of uncorrelated foregrounds and systematics, and joint constraints that significantly exceed the power of any single survey.
We specifically consider tracers of the baryon cycle: \ciium for the star formation rate (SFR), HI 21 cm for atomic hydrogen (diffuse gas reservoirs), and the CO rotational ladder for molecular hydrogen (star formation fuel stores). In addition, we consider the cross-correlation of these lines with \halpha spectroscopic surveys, a high-fidelity tracer of high SFR galaxies. Typically, forecasts compute the theoretical auto- and cross-power spectra by analytically expressing the two-halo, one-halo, and shot noise contributions independently or by considering only a dominant subset of the three \citep{Furlanetto_2007, Lidz_2008, Schaan_2021, Padmanabhan_2019, Padmanabhan_2022}. We develop a methodology that skips these intermediate steps and the corresponding implicit assumptions. We compute our signals by applying empirical line emission models to subhalos in state-of-the-art magnetohydrodynamical (MHD) simulations, which directly encode all three contributions to the power spectra signal. 
This provides a self-consistent, assumption-minimized forecasting framework that captures nonlinear clustering, halo occupation statistics, line–luminosity scatter, and covariance structure between lines, without decomposing the signal into analytically separated components.

An additional consequence of the relatively low signal-to-noise of current LIM tracers is the widespread use of deep field scan strategies (geometries that are deep in the redshift direction and small in sky area) in current-generation surveys. These enable stronger constraints on the redshift evolution of line emission but limit sensitivity to larger spatial scales. This also results in a mismatch between the $k$ modes probed in the line of sight and transverse directions, with measured $k_\perp$ and $k_\parallel$ differing by over an order of magnitude for some experiments. Typical forecasting formalisms do not account for this $k$-space anisotropy. In this paper, we develop a formalism for anisotropic window functions and mode counting, which is tailored for the calculation of survey sensitivities for these deep-field ($10^{0-2}\deg^2$), medium-resolution ($R\sim10^{2-3}$), LIM surveys. 

This paper is organized as follows. In Section \ref{sec:surveys}, we detail the three classes/generations of LIM surveys for \ciiumns, CO, and HI 21 cm, as well as \halpha spectroscopic galaxy catalogs. In Section \ref{sec:power_spec}, we describe the MHD-simulation-based forward modeling of the three-dimensional theoretical power spectra signal for four tracers. We formalize our auto- and cross-spectrum sensitivity calculations in Section \ref{sec:variance}. We present our forecasts for these nine statistics in Section \ref{sec:results}, discuss these in the context of three generations of LIM in Section \ref{sec:discussion}, and conclude in Section \ref{sec:conclusion}. 
Throughout this paper, our power spectra are presented in non-comoving units, with $k$ in Mpc$^{-1}$, line intensities in Jy/sr,  and logarithmic binning in $k$-space of $\delta \log k \sim 0.3$. We assume a \texttt{Planck2018} cosmology \footnote{flat $\Lambda$CDM, $H_0 = 67.66$ km/(Mpc s), $\Omega_{m,0} = 0.30966, T_\text{CMB, 0} = 2.7255~$K, $n_\text{eff} = 3.046, \Omega_{b,0} = 0.04897$} following \cite{2020A&A...641A...6P} \footnote{specifically, the ``TT, TE, EE + lowE + lensing + BAO" parameters from Table 2 of \cite{2020A&A...641A...6P}} and a Chabrier initial mass function (IMF) following \cite{2003Chabrier}. 

\section{Line Intensity Mapping Surveys \label{sec:surveys}}

\begin{table}[tp]
\centering
\begin{tabular}{c||c|c|c|c}
Tracer/Line & $\nu_\text{rest}$ (GHz) & $z$ & 
Instrument & $\nu_\text{ins}$ (GHz) \\
\hline
\hline
\ciium & 1901 & $0.52-1.01$ & 
TIM-SW & $1250-945$  \\
 & & $1.01-1.66$ & 
 TIM-LW & $945-714$ \\
\hline
HI 21 cm & 1.42 & $0.52-1.45$ & 
MeerKAT-UHF & $0.58-1.02$   \\
& & $0.52-1.66$ & SKA-Mid (Band 1) & $0.35-1.05$
\\
\hline
CO$_{J=4\rightarrow3}$ & 461 & $0.52-1.01$ & 
EoRSpec/FYST (LF) & $210 - 315$  \\
CO$_{J=5\rightarrow4}$ & 576 & $1.01-1.66$ & 
\\
\hline
\hline
spec-$z$ (\halphans) & 0.66$\mu$m & $0.52-1.66$ & 
NISP/\textit{Euclid}  & $0.92-1.85\mu$m  \\
\hline
\end{tabular}
\caption{An overview of tracers and corresponding instruments considered in our forecast; we examine three lines and galaxy redshifts in $z\sim0.5-1.7$. Frequencies for \ciiumns, HI 21 cm, and CO are listed in GHz; for \halphans, we instead list wavelengths in microns (1 $\mu$m $\equiv$ 299.79 THz). We also consider future variants of TIM, MeerKAT (in the form of SKA-Mid), and EoRSpec, as well as extensions to the currently planned survey with \textit{Euclid}, as next-generation surveys. The frequencies listed above are, respectively, the rest-frame emission frequency and the coverage of the instrument considered. MeerKAT does not go low enough in its frequency coverage, missing a portion of the highest redshift bin we consider; we only present hypothetical forecasts in this bin at $z\sim 1.5$ for a modified MeerKAT (see Section \ref{subsec:meerkat}).
}
\label{tab:lines}
\end{table}

We describe surveys of three different emission lines, one each in the far-infrared, submillimeter, and radio bands, as well as a galaxy spectroscopic redshift survey in the near-infrared; see Table \ref{tab:lines} for an overview. While the corresponding instruments are varied in implementation, design, and systematics, encompassing ground-, suborbital-, and space-based telescopes, they all currently plan to map the same deep field in the Southern sky: GOODS-South (or E-CDFS) roughly centered at RA $3\text{h}32\text{m}28\text{s}$, Dec $-27^\circ 48' 30''$ (J2000). 

For each line tracer, we define three generations of surveys: \sonens, \sfourns, and \shundns, mapping $1\deg^2$, $4\deg^2$, and $100\deg^2$ respectively. We set these to roughly correspond to the current generation of instrumentation that is already on-sky or imminent, the next generation of surveys that are a proposed reasonable upgrade or currently in development, and futuristic, forward-looking instrumentation that represents a reach goal over the next few decades in each class.

We calculate and present forecasts in 4 redshift bins over $z \sim 0.52 - 1.66$; see Table \ref{tab:lines}. The distinction in instrumental properties (e.g., noise and pixel counts) between the two modules of the Terahertz Intensity Mapper (TIM) offers a boundary for binning in $z$; we further divide each of its short wavelength (SW) and long wavelength (LW) $z$ ranges into two. While EoRSpec will be able to detect other CO lines, we select lines within the band of their LF module due to its superior expected noise performance.

\subsection{\texorpdfstring{\ciiumns}{[CII]-158um}: Terahertz Intensity Mapper (TIM) and future variants \label{subsec:tim}}
Our \cii \sone survey is the Terahertz Intensity Mapper (TIM; \cite{vieira2020terahertzintensitymappertim}), a balloon-borne far-infrared spectrometer set to launch in December 2027, which will survey a $1\deg^2$ field for 200 hours from 714 GHz (420 $\mu$m) to 1250 GHz (240 $\mu$m). The instrument consists of a 2-meter ambient-temperature primary mirror onboard a NASA Antarctic Long Duration Balloon (LDB). TIM will make a $R\sim250$ map of \ciium at $z\sim0.5-1.7$, with two distinct frequency modules - SW ($N_\text{spaxels} \sim$ 51) and LW ($N_\text{spaxels} \sim$ 51) - each covering roughly half the redshift (or wavelength) range.

TIM's instrumental noise is dominated by its warm ($\sim250~$K; passively cooled) primary and secondary mirrors and $\gtrsim160$~K optical filters, which are $\sim2$ orders of magnitude brighter within the observing band than the \ciium signal. 
We conceptualize both our \ciins-LIM \sfour and \shund instruments as utilizing a primary aperture cryogenically cooled to $\lesssim10$~K. 
Our \sfour survey, CryoTIM, is a hypothetical balloon-borne spectrometer like TIM, on the same 200-hour LDB flight, but with an actively cooled primary mirror, such that its noise is dominated by the order-of-magnitude fainter stratospheric background at LDB altitudes ($\sim 35$~km). For feasibility, we scale down the aperture to 0.5 m. Our \shund survey (SpaceTIM) is a probe-class 2-m space observatory at L2 (similar to PRIMA \citep{glenn2025prima}), limited by galactic and zodiacal light, achieving detector loading and sensitivity comparable to those assumed for \textit{Origins} Survey Spectrometer \citep{2021JATIS...7a1017B}, with $8\%$ emissive optics and $R\sim 250$. 
For simplicity, we assume the same bands and pixel counts for these experiments as TIM; the improvement in instantaneous sensitivity per detector is the dominant effect. Key properties are highlighted in Table \ref{tab:ciisurveys}.

\begin{table}[t]
\centering
\begin{tabular}{c|c||c|c|c|c|c}
Class & Instrument & $D$ & Band & NEI ($M\text{Jy}/\text{sr}\sqrt{\text{s}}$) & $T$ (hrs) & $A_\text{survey} \; (\deg^2)$  \\
\hline
\sone & TIM & 2~m & SW & 124.1 & 200 & $1\times 1$  \\
 & & & LW &  68.1 & &  $=1$  \\
\hline
\sfour & CryoTIM & 0.5~m & SW & 7.9 & 200 & $2\times 2$\\
 & & & LW & 3.2 & & $=4$ \\
\hline
\shund & SpaceTIM & 2~m & SW & .17 & 1000 & $10\times 10$ \\
 & & & LW & .13 &&  $=100$ \\
\hline
\end{tabular}
\caption{\label{tab:ciisurveys} List of \ciium experiments with corresponding parameters. The noise equivalent intensities (NEIs) of these three mapping surveys are (respectively) limited by warm primary optics, atmospheric foregrounds at stratospheric balloon altitudes, and galactic zodiacal light for a space observatory at L2. For simplicity, we assume the same detector number counts, bands, and spectral resolution for each of the three classes of far-infrared spectrometers. Refer to Table \ref{tab:lines} for definitions of the SW and LW bands.}
\end{table}

\subsection{HI 21 cm: MeerKAT and Square Kilometer Array-Mid \label{subsec:meerkat}}

The MeerKAT radio interferometric array is a 64-dish precursor to the mid-frequency Square Kilometer Array. LADUMA (Looking At the Distant Universe with the MeerKAT Array; \cite{Holwerda_2011, kazemimoridani2024lookingdistantuniversemeerkat}) is an ongoing single-pointing $\sim 3424$-hour survey in E-CDFS, tracing the 21 cm emission at $z\lesssim1.4$. 
For our \sone and \sfour surveys, we assume 1000-hour and 3000-hour (respectively) single-pointing observations in the MeerKAT UHF band ($580-1015$ MHz), at wideband coarse resolution ($\delta\nu = 132.812~k$Hz) with dish diameter of 13.5 m and system temperature $T_\text{sys} = 28~$K.
The single pointing field of view ranges from $\sim1-5\deg^2$ over $z=0.5-1.7$. Thus, at certain redshifts, the field of view (FOV) is larger than the cross-correlation coverage we are utilizing. We scale our observing time by this ratio of FOV to $A_\text{survey}$ in our sensitivity calculation.

The frequency coverage of MeerKAT UHF band only extends to $z\sim1.45$ for the 21 cm line, missing a significant portion of the highest redshift bin we consider. The MeerKAT sensitivity curves for this redshift bin are presented only for context, conceptualizing a MeerKAT 64-antenna configuration, fitted with the next-generation SKA-Mid Band 1 receivers.

For our \shund survey, we extend this configuration to a 5000-hour 100 $\deg^2$ survey with the proposed 197-dish SKA-Mid, including the MeerKAT antennas \citep{2022JATIS...8a1021S}, which also extends the frequency range down to 350 MHz, sensitive to the 21 cm line up to $z\sim3$.

\subsection{CO ladder: Epoch of Reionization Spectrometer/FYST and future extensions \label{subsec:eorspec}}

The Epoch of Reionization Spectrometer (EoRSpec; \cite{10.1117/12.2629338, huber2022ccatprimeopticaldesignepoch, Freundt_2024}) is a sub-millimeter $R\sim100$ spectrometer on the 6-meter Fred Young Submillimeter Telescope (FYST) \cite{CCAT_Prime_Collaboration_2022} in the high-altitude Atacama desert. As part of the CCAT Deep Spectroscopic Survey (DSS), EoRSpec will map two $2 \deg \times\ 2 \deg$ fields (E-CDFS and COSMOS) for 2000 hours each, at $210-420$ GHz. While the primary scientific goal of the CCAT DSS is to place limits on \ciium at $z\sim 4-7$, CO lines are also bright foregrounds at these wavelengths. Specifically, within the lower noise Low Frequency band \footnote{LF; $210-315$ GHz; NEFD (per beam, in median weather) $\sim81~m\text{Jy}\sqrt{\text{s}}/\text{beam}$; 
$N_\text{spaxels} \sim$ 6912, 80\% expected detector yield, unpolarized measurement mode} \cite{CCAT_Prime_Collaboration_2022}, the CO$_{J=4\rightarrow3}$ and CO$_{J=5\rightarrow4}$ transitions will be mapped from emitters at $z\sim0.5-1$ and $z\sim1-1.7$ respectively. We consider the intensity maps of these two lines for the SW and LW redshift ranges of our \ciium experiments. As EoRSpec uses a Fabry–Pérot interferometer (FPI) as its spectral element, we scale down the integration times of our survey by the ratio of \text{total bandwidth} to the \text{spectral resolution element}, a factor of $\sim 42$, given $\delta \nu \sim 2.6$ GHz at $R\sim 100$.

\begin{table}[t]
\centering
\begin{tabular}{c|c||c|c|c|c|c}
Class & Instrument & Spec.-Hours & $N_\text{beams}$ & FPI loss & $T$ (hrs) & $A_\text{survey} \; (\deg^2)$  \\
\hline
\sone & DSS-EoRSpec & $7\times10^4$ & 6912 & 1/42 & 500 & $1\times 1 = 1$   \\
\hline
\sfour & 7OT-EoRSpec & $9 \times 10^5$ & 24192 & 1/42 & 2000 & $2\times 2 = 4$\\
\hline
\shund & OCS-EoRSpec & $1\times 10^7$ & 8400 & 1 & 2000 & $10\times 10 = 100$\\
\hline
\end{tabular}
\caption{\label{tab:cosurveys} List of CO millimeter-wave experiments with corresponding parameters. We assume the same bands, detector yield ($80\%$) and spectral resolution for each of the three classes. The $1\deg^2$ and $4\deg^2$ surveys use an FPI, resulting in a loss in integration time of a factor of $\sim 42$, while the \shund survey uses on-chip spectrometers, with each beam being an independent spectrometer on sky. }
\end{table}

Our \sone survey adopts the DSS depth with EoRSpec: 500 hours over a $1\deg^2$ field.
In order to conceptualize next-generation surveys, we consider \cite{karkare2022snowmass2021cosmicfrontier}, which outlines a suggested staged experimental program to improve millimeter-wave LIM sensitivity. Ground-based detectors are already background-limited; hence, the necessary path forward is to increase the integrated spectrometer-hours on-field of millimeter telescopes. 
For the on-the-horizon \sfour survey, we consider the EoRSpec-like detectors populating the entire Prime-Cam focal plane on FYST, i.e., seven instead of two optics tubes (``\texttt{7OT}"), resulting in a $3.5\times$ increase in spectrometer beams on sky. We maintain the DSS survey strategy: 2000 hours on a $4\deg^2$ field. 

For our \shund survey, we hypothesize observing a $100\deg^2$ field for 2000 hours, with the same focal plane area as Prime-Cam, but filled with on-chip spectrometers (``\texttt{OCS}"). We consider each of the three detector modules per optics tube to hold $\sim400$ spectrometers (as suggested by \cite{karkare2022snowmass2021cosmicfrontier}), with no observation time loss due to the FPI scanning. These three survey classes correspond to a staged increase in spectrometer-hours by roughly an order of magnitude in each generation, similar to the staged program in \cite{karkare2022snowmass2021cosmicfrontier}; see Table \ref{tab:cosurveys}.

\subsection{Galaxy counts with \texorpdfstring{\halphans}{Halpha} spec-\texorpdfstring{$z$}{z}'s: \textit{Euclid} Deep Field Spectroscopy}

We consider the \textit{Euclid} Deep Field Spectroscopic Survey \cite{EuclidOverview, euclidcollaboration2025euclidquickdatarelease, 2025EuclidDAWN, 2022A&A...662A.112E, 2022A&A...662A..92E} from the Near-Infrared Spectrometer and Photometer (NISP; \cite{2025Euclid}), specifically the 10~$\deg^2$ coverage in the Fornax field 
(which overlaps with E-CDFS) as both our \sone and \sfour survey. 
Our \shund survey is a hypothetical extension of \textit{Euclid}'s spectroscopic galaxy survey with \halpha as the redshift tracer,
i.e., we assume that a next-generation survey can achieve the flux limit of EDF-F over a contiguous $10\deg\times10\deg$ footprint.

\section{Power Spectrum Simulation \label{sec:power_spec}}

There are two key ingredients for computing the sensitivity for a given LIM survey: the theoretical power spectrum signal and the expected instrumental noise, both as functions of redshift $z$ and mode $k$. 
In this Section, we detail the calculation of the three-dimensional distribution of line emissions at four redshifts by applying empirical models to volumes from a magnetohydrodynamical (MHD) simulation. This general framework is described in Section \ref{subsec:simim}, with the next four subsections describing empirical models for the four tracers. 
Taking the auto- and cross-spectra of these distributions allows us to write down the theoretical signal that the experiments in Section \ref{sec:surveys} will measure. In Section \ref{sec:variance}, we write the formalism for calculating the signal-to-noise ratio (S/N) in a given survey's measurement of the true power spectra signal, including counting $k$ modes for anisotropic surveys and window functions due to a survey's finite coverage and resolution.

\subsection{Painting magnetohydrodynamical simulations with empirical models of line emission \label{subsec:simim}}

\begin{figure*}[tbp]
\centering
\includegraphics[width=0.49\linewidth]{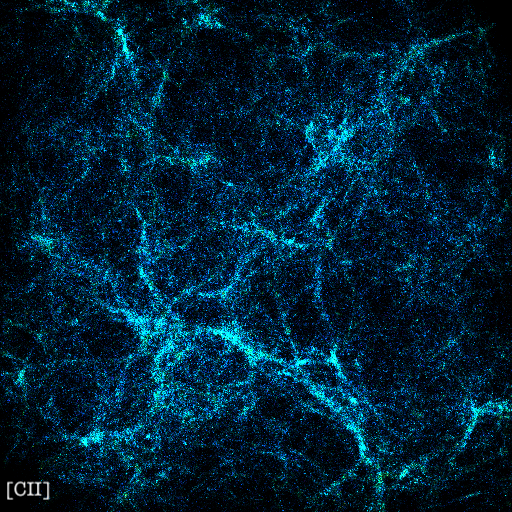}
\includegraphics[width=0.49\linewidth]{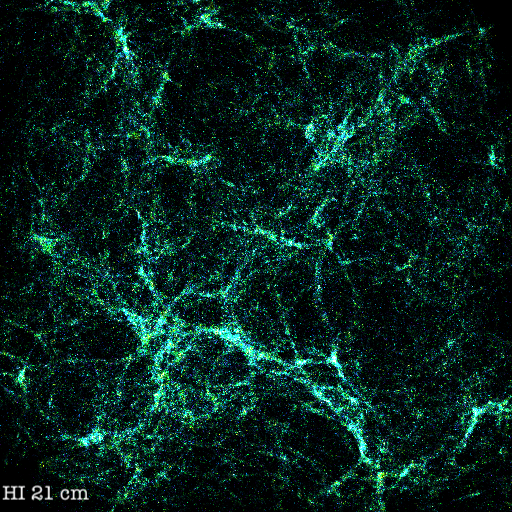}
\includegraphics[width=0.49\linewidth]{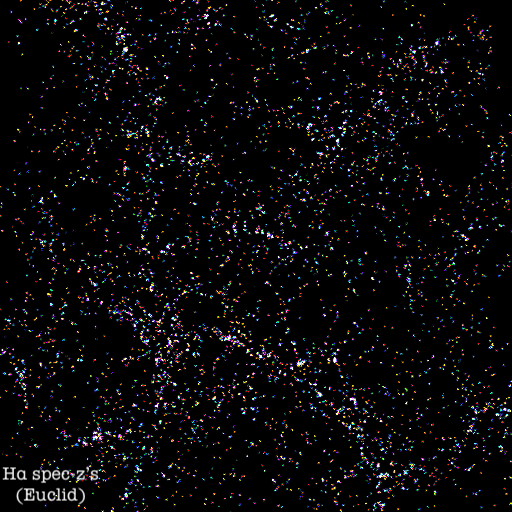}
\includegraphics[width=0.49\linewidth]{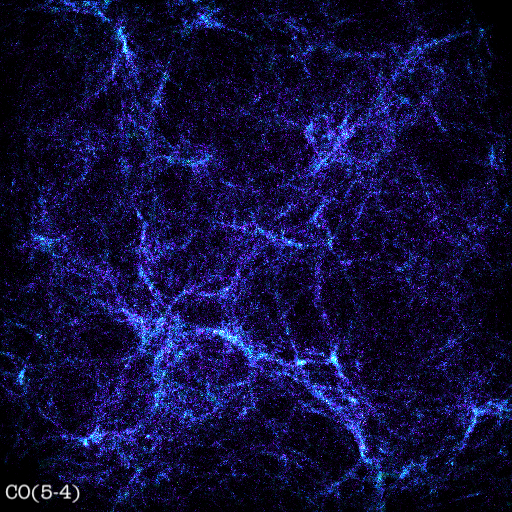}
\caption{\label{fig:visualization}
Constructed three-dimensional distributions at $z\sim1.17$ of (\textit{clockwise from top left}) line emission from \ciiumns, HI 21 cm, CO$_{5\rightarrow4}$, and galaxy densities from a \textit{Euclid}-depth \halpha spectroscopic redshift survey. The entire \texttt{TNG100} volume (with side length 75 Mpc/$h$) is shown; all species trace the same cosmic web at large scales, but exhibit variations at small scales dictated by the empirical emission models dependent on subhalo properties and relevant correlation coefficients (see Figure \ref{fig:ccc}). We made the visualizations with the \texttt{yt} Python toolkit \citep{2011ApJS..192....9T}.}
\end{figure*}

Past works \citep{Furlanetto_2007, Lidz_2008, Schaan_2021, Padmanabhan_2019, Padmanabhan_2022} typically specify analytical expressions for the power spectrum of LIM tracers, decomposing them into the two-halo and one-halo clustering contribution, as well as a Poisson-dominated shot-noise term,
\begin{align}
\label{eqn:analytic}
P_{ij}^t = P_{ij}^\text{2-halo} + P_{ij}^\text{1-halo} + P_{ij}^\text{shot},
\end{align}
for tracers $i$ and $j$. Typically, computing the two-halo contribution requires a model for the underlying dark matter power spectrum, as well as estimates of two linear biases and mean intensities, the latter of which subsequently depend on models for the luminosity function (LF) and the halo mass function (HMF). Similarly, computing the one-halo term involves assuming a spatial mass distribution profile (e.g., Navarro-Frenk-White \cite{1997ApJ...490..493N}) for single dark matter halos, along with models for the line LF and HMF. The shot term depends on the multi-line LF. Computing theoretical signals for the auto- and cross-power spectra using this class of methodology is non-trivial, depends on model choices for LFs, HMFs, halo profiles, and biases, and incorporates the underlying assumptions encoded in those models. 

As an alternative to attempting to model the $k$-space power spectra directly, in this section, we detail our methodology for forward-modeling three-dimensional distributions of line intensities and emitter point clouds from magnetohydrodynamic (MHD) simulation volumes. These simulations directly encode redshift evolution of the underlying three-dimensional dark matter density field and large-scale structure, galaxy formation and feedback, and extragalactic physics (including LFs, HMFs, and halo profiles), allowing us to skip computing intermediate quantities or models and directly incorporating the physics encoded in the MHD simulations. 

We use the open-source Python utility SimIM\footnote{\href{https://simim.readthedocs.io/}{simim.readthedocs.io}} \footnote{\href{https://github.com/shubhagrawal30/simim\_public}{github.com/shubhagrawal30/simim\_public}
} 
\footnote{Implementation code available at \href{https://github.com/shubhagrawal30/crossLIM}{github.com/shubhagrawal30/crossLIM}}, for modeling the extragalactic sky in radio, sub-millimeter, and far-infrared.
This formalism has been previously used in studies of the aggregate CO line emission at high redshifts \citep{Li_2016, Chung_2019, 2020ApJ...901..141K, Keenan_2020, Keenan_2022}.
We present an overview of the methodology as it applies to modeling theoretical power spectra of line emission.

\begin{figure*}[tbp]
\centering
\includegraphics[width=\linewidth]{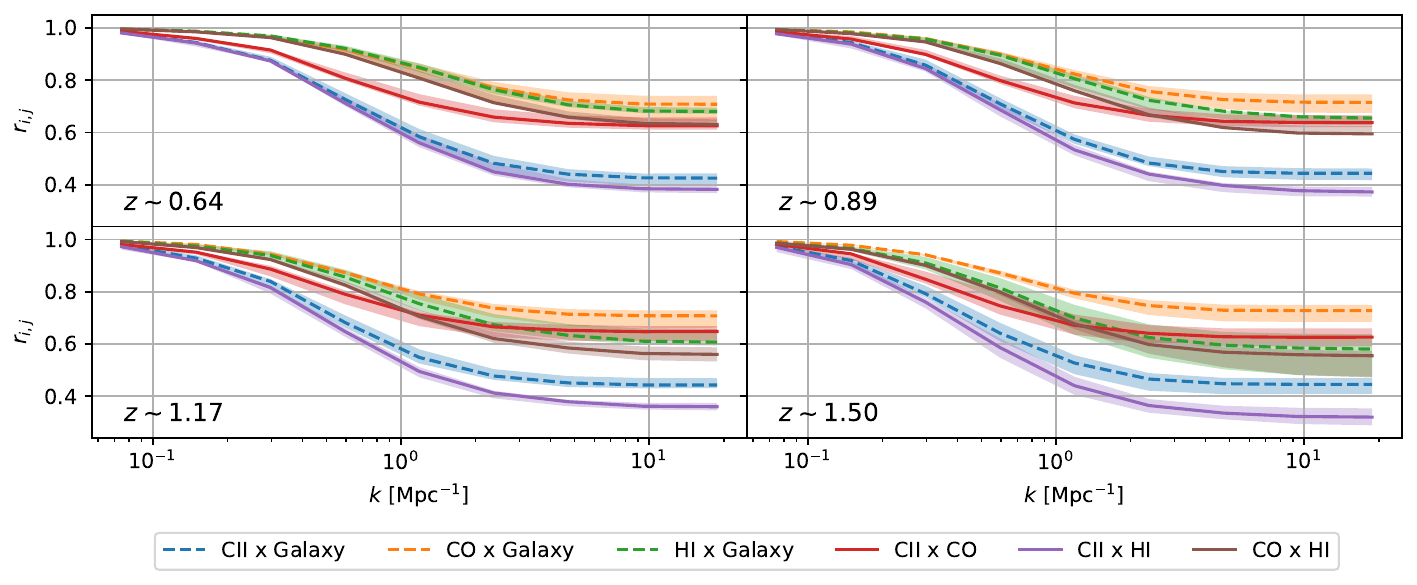}
\caption{
\label{fig:ccc}
Cross-correlation coefficients, $r_{ij}(\bar k, z) = P_\text{cross, $i\times j$} / \sqrt{P_\text{auto, $i$} P_\text{auto, $j$}}$, for all six possible pairs for the four tracers (\ciiumns, HI 21 cm, CO, \halphans-galaxies) considered, in our four $z$ bins. These coefficients encapsulate new information present in the cross-correlation, unavailable from measurements of just the two auto-correlations; the values start to deviate from unity as we progress from the two-halo to the one-halo and shot regimes at higher $k$'s. 
Different tracers have different cross-shot behaviors; these also evolve with $z$, indicating new astrophysical potential from tomographically measuring cross-power spectra.
The cross correlation coefficients encode differences in scaling relations of line luminosities to galaxy properties, intrinsic scatter in line luminosities, and line-line covariance, and can inform models of galaxy evolution and the baryon cycle. } 
\end{figure*}

We start with an MHD simulation - the IllustrisTNG suite \citep{2018MNRAS.477.1206N, 2018MNRAS.475..676S, 2018MNRAS.475..648P, 2018MNRAS.475..624N, 2018MNRAS.480.5113M}, specifically the TNG-100-1 cube - with side length $75 ~\text{Mpc}/h$ and $1820^3$ initial gas cells and dark matter particles, which is adequate for exploring cross-correlations at $k\gtrsim10^{-1}$ Mpc$^{-1}$. This simulation suite is provided with a subhalo catalog constructed using SUBFIND \citep{2001MNRAS.328..726S, 2009MNRAS.399..497D}, identifying three-dimensional position vectors for each subhalo as well as salient properties, e.g., total or stellar mass.
Because these subhalos are identified in the full-MHD simulation, their positions, masses, merger histories, and survival incorporate the back-reaction of baryonic physics on the underlying matter field, and therefore differ from those in a dark-matter-only $N$-body realization.
We apply sub-grid line emission physics models (described in the next four subsections), painting each subhalo with line intensities, brightness temperatures, or booleans indicating detectability with a galaxy count survey. 
This 
yields, for each considered redshift and tracer, a three-dimensional distribution of intensities (or points for the galaxy survey).
These are gridded at a physical scale much smaller than the beam of each survey ($\sim 0.25$ Mpc), point clouds within each voxel are coadded, and a Fourier transform yields the three-dimensional $k$-space distribution. Under isotropy in the TNG100 simulation volume, we average in spherical shells with $d
\log_{10}(k) = 0.3$ to get auto- and cross-spectra predictions. 

In Sections \ref{subsec:CII}$-$\ref{subsec:halpha}, we detail the empirical line emission models used to construct these three-dimensional intensity distributions of \ciiumns, CO, \halphans, and HI 21 cm. 
For the first three tracers, we use scaling correlations with subhalo star formation rates, calibrated by observations of sources at relevant redshifts; for 21 cm, we utilize a HI-halo mass relation. We apply these empirical models to sub-halos to encode the structure within halos and model the one-halo clustering term. 
These outputs are visualized in Figure \ref{fig:visualization}; all four trace the same cosmic web in \texttt{TNG100} at large scales, but deviate at small scales, due to galaxy and halo-scale astrophysics.
The relations of line luminosities to SFR or subhalo mass are non-linear and distinct, resulting in distinct behaviors in the cross-correlation coefficients $r_{ij}(\bar k, z)$. These $r_{ij}$ are also dependent on both $k$ and $z$ as shown in Figure \ref{fig:ccc}. The cross-correlation coefficients encode astrophysical information not accessible with auto-correlations alone, including bulk covariance and intrinsic scatter in line luminosities. Note that this small-scale information is directly encoded in our forward models from the three-dimensional MHD simulation and does not require an analytical formulation for the cross-shot terms in the power spectra.

We consider two classes of stochasticity in constructing these three-dimensional distributions of line luminosities: intrinsic log-normal scatter in source properties \cite{2013Behroozi, delooze+14}, as well as uncertainties in fitted parameters in the empirical line models. We repaint the entire TNG100 volume $N=50$ times to get three-dimensional Monte Carlo (MC)  realizations, sampling the model space of line intensity distributions based on galaxy properties. For the complete TNG100 galaxy catalog at a redshift, for each of the MC realizations, we set an independent set of model parameters given the uncertainties in the empirical calibration fits for SFR, HI mass, or line intensities. We additionally add log-normal scatter to the source property to model for intrinsic variability and scatter in sources with the same halo properties. These models and property scatters are based on empirical measurements, as noted in the following subsections. For each of the MC samples of the three-dimensional volumes, we independently grid and compute power spectra to obtain expected scatters in the signal due to modeling uncertainty.

\subsection{The \texorpdfstring{\ciiumns}{[CII]-158um} fine structure line \label{subsec:CII}}

Observational constraints on the \ciium luminosity function exist in the local $z\sim 0$ universe, from 
\textit{Herschel}/PACS observations of $\sim500$ luminous infrared galaxies \citep{2017ApJ...834...36H}, 
and at $z\sim4-6$ by the ALPINE-ALMA survey \citep{2020ApJ...905..147Y, 2021A&A...646A..76L}.
As cosmic star formation peaks at $z\sim1-2$ \citep{Madau_2014, Walter_2020} and the excess UV photon flux warms dusty halos around star-forming regions,
the mean \ciium intensity is expected to become dramatically brighter, and then fall by order(s) of magnitude to the present day. 
Two competing groups of models have been proposed to quantify this cosmic evolution history of \ciiumns. The first class uses SFR$-L_\text{\ciins}$ relations,
calibrated through source studies \cite{delooze+14} or semi-analytical simulations \citep{2018A&A...609A.130L, 2019ApJ...882..137P, bethermin+22, Yang_2022}, which find a roughly linear relationship between galaxy SFR and $L_\text{\ciins}$, that does not
show a strong redshift evolution. The second class suggests \ciium luminosity to be proportional to carbon mass in halos and are parameterized by mean electron kinetic temperatures; these predict much higher \ciium intensities at cosmic noon \citep{2012Gong, Pullen_2018}.
These have also been calibrated to the tentative detection of bright diffuse \ciium emission at $z\sim2.6$ at large scales with all-sky \textit{Planck} HFI maps \citep{Pullen_2018, Yang_2019}.
Recent results from a search for sky-averaged \ciium emission in \textit{Herschel}/SPIRE maps of the COSMOS deep field across cosmic noon \citep{agrawal2025farinfraredlineshiddenarchival} 
seem to instead exclude the latter at high significance ($\gtrsim3\sigma$) and signal a preference for the former. 
  
For this forecast, we conservatively follow one of the lower intensity star-formation-tracing models of the \ciium luminosity functions at $\sim0.5-1.7$; 
the \cii signal could potentially be up to an order of magnitude brighter at these redshifts, resulting in a corresponding increase in the S/Ns we calculate.
Starting with masses in our MHD subhalo catalog, we use the empirical \cite{2013ApJ...770...57B} SFR-halo mass relation to assign each source a star formation rate.
We apply a 0.3 dex log-normal scatter to model intrinsic variations as per \cite{2013ApJ...770...57B}. We use this empirical scaling relation instead of the SFRs provided in the TNG subhalo catalog, as we expect the former to be more robust and based on observations at relevant redshifts \cite{Vogelsberger_2014, 2015MNRAS.447.3548S, 2018MNRAS.473.4077P, 2019MNRAS.485.4817D}.

The constraints from \cite{delooze+14} on scaling between \ciium luminosity and SFR from sources at $z\sim0-6$ are then used to get $L_{\text{\cii}}$ for each subhalo:
$\log_{10} \text{SFR}_{M_\odot/\text{yr}} = m \log_{10} L_{\text{\cii}, L_\odot} + b $, with $m=1.01\pm 0.02$, $b=-6.99\pm0.14$, and an intrinsic log normal scatter of 0.42 dex \citep{delooze+14}.

\subsection{21 cm from neutral hydrogen \label{subsec:HI}}

\cite{Padmanabhan_2016, Padmanabhan_2017} develop a halo model for neutral hydrogen, placing constraints on the HI-halo mass relation and the HI radial density profile using observations of HI in local galaxies, intensity mapping at $z\sim 2$, and UV-O Damped Lyman Alpha (DLA) systems at $z\sim5$. The average HI mass in a halo of dark matter mass $M$ is modeled as:
\begin{align}
M_{\text{HI}}(M,z) = \alpha f_{\text{H}} &M 
\left( \frac{M}{10^{11}\,h^{-1}M_{\odot}} \right)^{\beta}
\times
\exp \left[ - \left( \frac{v_{c}}{v_c(M,z)} \right)^3 \right],
\end{align}
with cosmological hydrogen fraction $f_H = (1 - Y_{\mathrm{He}})\;{\Omega_b}/{\Omega_m}$, primordial helium fraction $Y_\text{He}\sim 0.24$, and the halo circular velocity $v_c(M,z) = \sqrt{{GM}/{R_{\mathrm{v}}(M,z)}}$ defined for the virial radius, 

\begin{align}
\frac{R_{\mathrm{v}}(M,z)}{46.1 \, \mathrm{kpc}} =  
\left( \frac{\Delta_v \Omega_m h^2}{24.4} \right)^{-\frac{1}{3}}
\left( \frac{1+z}{3.3} \right)^{-1} 
\left( \frac{M}{10^{11} M_{\odot}} \right)^{\frac{1}{3}},
\end{align}
for the virial overdensity parameter $\Delta_v(z) = 18\pi^2 + 82(\Omega_m(z)-1) - 39(\Omega_m(z)-1)^2$ \citep{Bryan_1998}. The HI mass is parameterized by an overall normalization $\alpha$, a logarithmic scaling coefficient with halo mass $\beta$, and a lower virial velocity cutoff at $v_c$; these are constrained by the observational compilation to be $\alpha\sim0.09\pm 0.01, \beta\sim -0.58\pm0.06$, and $\log_{10}(v_c/1~\text{km/s}) \sim 1.56\pm 0.04$. These parameters have been shown to be consistent \citep{padmanabhan2024hiintensitymappingpower} with the recent tentative detection of the HI 21 cm auto-power spectrum at $z\sim 0.32, 0.44$ at small scales with MeerKAT \citep{paul2023detectionneutralhydrogenintensity}.

Under the assumption of high enough spin temperatures ($T_s > T_* \sim 0.0682$ K), the upper hyperfine occupation fraction is given as $f_1\sim 3/4$ \citep{2013tra..book.....W}. Thus, we can obtain the 21 cm line luminosity as:
\begin{align}
L_\text{21} = f_1 \; (M_\text{HI}/ m_H) \; A_{10} \; h \nu_{21, \text{rest}},
\end{align} 
with the Einstein spontaneous emission
coefficient $A_{10}\sim 2.869\times10^{-15}~\text{s}^{-1}$ \citep{2013tra..book.....W}. 

\subsection{CO rotational ladder \label{subsec:CO}}
There are a multitude of models for the redshift evolution of the aggregate or $J=0$ CO line intensity and its dependence on (sub)halo mass \citep{2016ApJ...830...34K, Li_2016, 2013ApJ...768...15P, Lidz_2011}. However, 
because we aim to model specific $J+1\rightarrow J$ transitions of the CO ladder ($J=3$ and $J=4$), we require models that specify particular high-$J$ lines \citep{2014ApJ...794..142G,2015ApJ...810L..14L,2016ApJ...829...93K}. We use \cite{2016ApJ...829...93K}, which presents 
constraints on the $L_\text{CO}/L_\text{FIR}$ relations from archival Herschel SPIRE Fourier Transform Spectrometer spectra:
\begin{align}
\log L_\text{FIR} = a_J \log L_{\text{CO}(J+1\rightarrow J)} +b_J,
\end{align}
with $a_3=1.09\pm0.05, b_3=1.2\pm0.4, a_4=1.05\pm0.03, b_4=1.8\pm0.3$, for the full sample ($n=108$ and $n=195$ fits for $J=3$ and $J=4$ respectively). We estimate $L_\text{FIR}$ from the SFR, assuming a Chabrier IMF and $L_\text{FIR}\sim 0.6 L_\text{IR}$ \citep{2003PASP..115..763C}. The SFRs are calibrated to the subhalo mass and redshift, as in Section \ref{subsec:CII} with the \cite{2013ApJ...770...57B} empirical relation.

\subsection{Spectroscopic galaxy redshift surveys with \texorpdfstring{\halphans}{Halpha} \label{subsec:halpha}}
We take SFRs and line luminosities of \halpha to follow the fiducial \cite{1998ARA&A..36..189K} linear relation:
\begin{align}
L_{\text{\halphans}, \text{erg/s}} = 1.26 \times 10^{41} \; \times \; \text{SFR}_{M_\odot/\text{yr}},
\end{align}
and add a 0.3 dex log-normal intrinsic luminosity scatter. 
This calibration has been demonstrated at $z\sim2$ \citep{Shivaei_2015}. While the linear relation can be insufficient, especially for dwarf galaxies \citep{2007ApJ...671.1550P}, this happens at low SFR/\halpha luminosities, i.e., for sources which would not be detected by our galaxy surveys and hence do not need to be modeled for our analysis.
Given the line flux detection limit for \textit{Euclid}, we tag each source as detectable or undetectable in our three-dimensional volumes at each redshift. Simulations have shown that the flux limit of \textit{Euclid} in its deep fields for $3.5\sigma$ detections is $6.9\pm2.8\times 10^{-17}$ erg s$^{-1}$ cm$^{-2}$ \citep{2023EuclidXXX}.
After gridding, rather than coadding line intensities, we count the number of galaxies tagged as detectable by \textit{Euclid} in each cell. This yields a three-dimensional distribution of galaxy over- and under-densities after a mean subtraction, allowing us to model the galaxy auto- and cross-power spectra.

At \textit{Euclid} grism spectral resolution, the \halpha line is expected to be blended or confused with the nearby [NII] doublet \citep{Bagley_2020, Cagliari_2024}; we account for this by inflating the \halpha fluxes, assuming $L_\text{\halpha + [NII]} \sim (1+ 0.4) L_\text{\halpha}$ \citep{Faisst_2018}. The mean source number densities we obtain in our four redshift bins are comparable to past estimates for the \textit{Euclid} deep field spectroscopic catalogs \citep{Pozzetti_2016, 2025EuclidV}.

\begin{figure*}[thp]
\centering
\includegraphics[width=\linewidth]{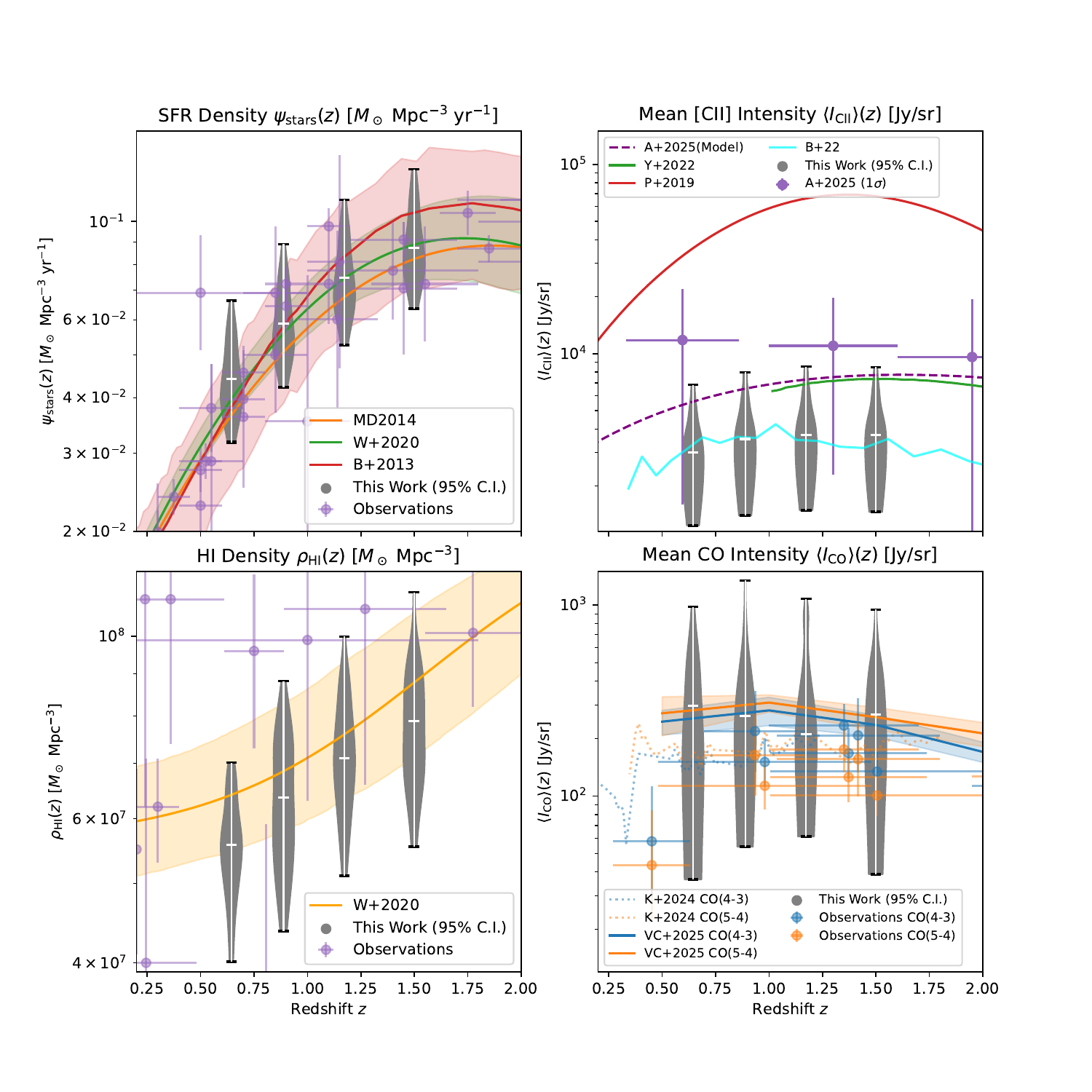}
\caption{We compare (\textit{clockwise from top left}) star formation rate density, sky-averaged mean \ciium intensity, sky-averaged mean CO intensity, and neutral hydrogen density, obtained from our MHD simulation-based forward modeling methodology, against literature models and measurements. These are described in detail in Section \ref{subsec:validation}: SFR measurements from UV and IR data are from the compilation in \cite{Madau_2014}, \ciium measurements are from \textit{Herschel} data \cite{agrawal2025farinfraredlineshiddenarchival}, HI density measurements from the compilation in \cite{Walter_2020}, and CO mean intensities derived from \cite{decarli+20,lenkic+20,boogaard+23,Keating_2020}. The scatter in the MC sampling ($N=50$) is represented as violin plots between the extrema, with the median value marked. Our models are in general agreement with 
the literature.}
\label{fig:validation}
\end{figure*}

\subsection{Validation against Observations and Models of Redshift Evolution \label{subsec:validation}}

In this subsection, we demonstrate that the amplitude and redshift evolution of four key quantities obtained from our MHD forward-model methodology is consistent with literature models and observations. Figure \ref{fig:validation} considers the star formation rate density (SFRD) $\psi_\text{stars}(z)$, neutral hydrogen density $\rho_\text{HI}(z)$, mean \ciium intensity $\langle I_{\nu, \text{\ciins}}\rangle(z)$, and mean CO  intensity $\langle I_{\nu, \text{CO}}\rangle(z)$ (for both $J=4$ and $J=3$) over $z\sim0.5-1.7$.
Our predictions, represented as violin plots, are in general agreement with literature models and observations, with a large scatter to encode uncertainty in current constraints.

\paragraph{Cosmic SFR}: \cite{Madau_2014} (\texttt{MD2014}) provides a compilation of SFR measurements from UV and IR data from $z\sim 0-8$, along with a best-fit model for the overall evolution of the cosmic SFR density. We also show similar models from \cite{Walter_2020} (\texttt{W+2020}) and \cite{2013ApJ...770...57B} (\texttt{B+2013}) with their associated 1$\sigma$ scatter. Note that there is a mismatch of the underlying IMF assumed by \cite{Madau_2014} and \cite{Walter_2020}, which assume the Salpeter IMF instead of the Chabrier. We scale their models by a factor of 0.63 \citep{Madau_2014} to compare consistently.

\paragraph{\ciium Mean Intensity}: We overplot recent measurements of the mean intensity of \ciium from archival \textit{Herschel}/SPIRE maps in the COSMOS deep field at $z\sim0.3-2.9$ \citep{agrawal2025farinfraredlineshiddenarchival} (\texttt{A+2025}), which are consistent with the evolution predicted by the SFR-scaling relation from \citep{bethermin+22} (\texttt{B+2025}) as well as the Santa Cruz Semi-Analytical Model (SAM) lightcone simulations from $N$-body simulations \citep{Yang_2019} (\texttt{Y+2022}). We include predictions from \cite{Padmanabhan_2019} (\texttt{P+2019}), which fits an empirical model for the \ciins-LF to the \cite{Pullen_2018, Yang_2019} measurement of \ciium from \textit{Planck} high-frequency maps; this is roughly an order of magnitude brighter than the SFR-tracing models and the \cite{agrawal2025farinfraredlineshiddenarchival} measurements. We also include the model from \cite{agrawal2025farinfraredlineshiddenarchival} which scales a SFR-tracing \ciium model by a best-fit constant amplitude to current constraints at $z\sim0-6$. As noted in Section \ref{subsec:CII}, we follow the pessimistic set of models that predict \ciium to be a few percent of the CIB at frequencies corresponding to these redshifts.

\paragraph{CO Mean Intensity}: We plot models for the mean intensities of CO$_{4\rightarrow3}$ and CO$_{5\rightarrow4}$ obtained from the Simulated Infrared Extragalactic Sky (SIDES) simulations \citep{bethermin+22, vancuyck2025unveilingevolutionexcitationladder} (\texttt{VC+2025}), and from mock tomographies of CO emitters constructed from the TNG300 simulation \citep{karoumpis2024ciilineintensitymapping} (\texttt{K+2024}). The predicted LF for both of these models for CO$_{4\rightarrow3}$ and CO$_{5\rightarrow4}$ are consistent with LF measurements from the ALMA Spectroscopic Survey (ASPECS) in the Hubble Ultra Deep Field at $z\sim1-2$ \citep{2020ApJ...902..110D}. We also include measurements of the integrated CO luminosity from blind CO deep fields \citep{decarli+20,lenkic+20,boogaard+23} and line intensity mapping experiments \citep{Keating_2020}. CO studies in this redshift range utilize primarily CO(2-1), CO(3-2) and CO(4-3), we convert the integrated intensities to CO(4-3) and CO(5-4) assuming the line ratios from \cite{boogaard+20} (for $z\sim1$ these are similar to the ratios reported by \cite{daddi+15}). 
The CO(5-4)/CO(2-1) and CO(4-3)/CO(2-1) ratios for a subset of galaxies in these samples has been measured directly in \cite{boogaard+20}, and the energy differences between e.g. the CO(4-3) and CO(3-2) line are smaller than the differences with the ground state, so the uncertainty in this conversion is smaller than in the conversion of any given line to CO(1-0) \citep{keenan+25b}.

\paragraph{HI Density}: We plot the 1-$\sigma$ empirical model constraint from \cite{Walter_2020} (\texttt{W+2020}) for $\rho_\text{HI}(z)$ over $z\sim0-4$, as well as the compilation of observations used therein. These include 21 cm measurements in the local universe, 21 cm stacking and LSS cross-correlation at intermediate redshifts, and measurements of damped Ly$\alpha$ systems at high redshifts.

\section{Instrumental Sensitivities to Power Spectra \label{sec:variance}}

\begin{figure*}[tb]
\centering
\includegraphics[width=\linewidth]{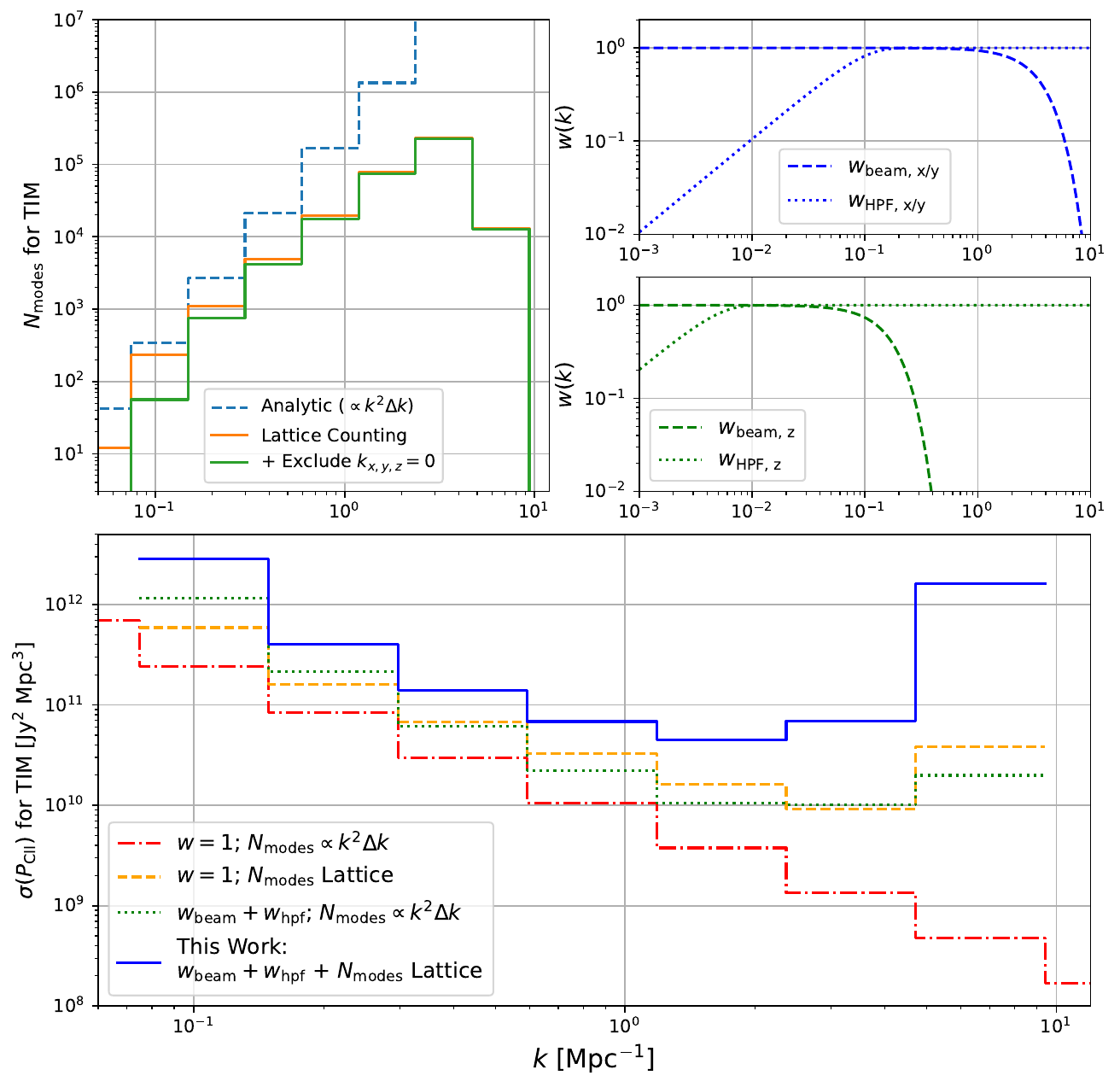}
\caption{\label{fig:windowmodes}
Key components in our calculation of a LIM survey's sensitivities to the power spectra. As a representative example, we present quantities for the Terahertz Intensity Mapper \ciium $1 \deg \times 1 \deg$ survey, in the third redshift bin at $z\sim 1.17$.
(\textit{Top Left}) Comparison of $k$-mode counting in our formalism to the fiducial analytical expressions, wherein $N_\text{modes}$ is proportional to the volume of thin isotropic shells in $k$-space, i.e., $N_\text{modes} \propto k^2 \Delta k$. Accounting for survey geometry anisotropy and potential contamination in zero-wavenumber modes ($k_{x, y, \text{or} z} = 0$), we conservatively directly bin and count modes from our $k$-space lattice and also exclude modes where any of the components $\bar k$ are zero ($k_x$, $k_y$, or $k_z$ = 0), resulting in lower values for $N_\text{modes}$, and a cutoff at lower $k$ magnitudes due to mismatch in the minimum modes measured in spatial and spectral directions.
(\textit{Top Right}) Two classes of window functions, $w_\text{beam}$ and $w_\text{hpf}$, due to the survey's non-zero resolution and finite coverage, attenuate high and low $k$ modes, respectively. We consider each class in all three Cartesian directions; the mismatch in the modes sampled in the spectral and spatial directions is apparent.
(\textit{Bottom}) Effect of applying the two main corrections (lattice-based mode counting with exclusion of modes with a vanishing Cartesian component, and window functions) in our analysis as per Eqn. \ref{eqn:noisecross} to the sensitivity of TIM to the \ciium auto-power spectrum.
}
\end{figure*}

\subsection{Finite spectral and spatial resolution and coverage}

A LIM survey (different tracers indexed by the subscripts $i, j$) attempts to measure the true real-valued intensity field signal $\delta_{i}^t(\bar{x})$, which is a biased tracer of the underlying dark matter density field, $\delta_{i}^t \propto b_i \delta_{i}^{DM}$, with some additive (assumed Gaussian) real-space noise $n_{i}$. Due to its finite coverage and resolution, a survey actually measures a windowed version of the true field \citep{Fronenberg_2024}:
\begin{align}
\label{eqn:window}
\tilde{\delta}_{i}^m(\bar{k}) &= \int_{\bar{k'}} \mathcal{W}_{i}(\bar k, \bar{k'}) \; \tilde\delta_{i}^t(\bar{k}) \; d\bar{k'} + \tilde{n}_{i}(\bar{k}) 
\\
&\sim w_{i}(\bar{k}) \tilde\delta_{i}^t(\bar{k}) + \tilde{n}_{i}(\bar{k}),
\end{align}
where we have approximated the transfer function $\mathcal{W}(\bar k, \bar{k'})$ as a simple multiplicative factor $w(\bar{k})$, similar to past works \citep{Li_2016, Bernal_2019, Padmanabhan_2022, Padmanabhan_2023}. $\bar x$ and $\bar k$ are real- and Fourier-space 3-vectors respectively; $\tilde\delta_{i}^m$ and $\tilde\delta_{i}^t$ are Fourier $k$-space analogs of the real-space $\delta_{i}^m$ and $\delta_{i}^t$. The superscripts $m$ and $t$ denote the measured and true power spectrum, respectively.  The power spectrum (a two-point statistic) is the ensemble average,
\begin{align}
\label{eqn:denfields}
(2\pi)^3 \;P(k) \; \delta_D^3(\bar k - \bar k') &= \langle \tilde\delta(\bar k) \tilde\delta^*(\bar k') \rangle.
\end{align}

We consider two contributions to $w_i(\bar k)$. High $k$-modes in each of the three directions (two spatial, one spectral; indexed by $m$) are attenuated by the (assumed Gaussian) beam of a survey:
\begin{align}
\label{eqn:beamwindow}
w_{i, \text{beam}}(\bar{k}) \sim 
\underbrace{\exp(-k_x^2 \sigma_{i,  x}^2 / 2) \; \exp(-k_y^2 \sigma_{i,  y}^2 / 2)}_\text{spatial beam} \;\;
\underbrace{\exp(-k_z^2 \sigma_{i,  z}^2 / 2)}_\text{spectral resolution}
\sim  \prod_m \exp(-k_m^2 \sigma_{i,  m}^2 / 2),
\end{align}
where the mode spread $\sigma_{i, m}$ is in distance units and defined differently for spatial and spectral directions,
\begin{align}
\sigma_{i, x/y} &= \chi({z}) \times \frac{\theta_{\text{FWHM}, i, x/y}}{\sqrt{8\ln 2}},  \\
\label{eqn:sigz}
\sigma_{i, \mathrm{z}} = \frac{d\chi}{d\nu} \delta\nu &= \frac{c}{H({z})} \frac{(1+{z})^2} {\nu_\text{rest}} \frac{\delta\nu_{\text{FWHM}}}{\sqrt{8\ln 2}};
\end{align}
$\theta_{\text{FWHM}}$ and $\delta \nu_{\text{FWHM}}$ are the angular full-width at half-maximum of the spatial beam and spectral resolution, respectively (Cartesian dimension is denoted as $\mathrm{z}$ to avoid confusion with the redshift $z$ in Eqn. \ref{eqn:sigz}). $\chi(z)$ and $H(z)$ are the comoving distance and Hubble parameter at redshift $z$ respectively.
We assume that the finite resolution of LIM surveys is much coarser than that of galaxy surveys, i.e., $w_{\text{LIM}, \text{beam}}w_{\text{gal}, \text{beam}}\sim w_{\text{LIM}, \text{beam}}$. 

Additionally, finite spectral and spatial coverage result in a finite $k$-space resolution. This is crucial to consider for LIM surveys, where astrophysical/terrestrial foreground contamination, cosmic background continuum, and time-domain noise (typically 1/$f$, or pink) corrupt modes close to $k_{i}\sim0$ in all three directions. Conservatively, we assume that the finite $k$-space resolution (and the corresponding {mode}-space point spread function) makes low-$k$ modes inaccessible; this is modeled using a high-pass filter in each of the three dimensions, assuming a survey footprint with Gaussian tapering at the edges \citep{Bernal_2019, Padmanabhan_2022, Padmanabhan_2023}:
\begin{align}
w_{i, \text{hpf}}(\bar{k})^2 \sim \prod_m \big[1 - \exp\big( - k_m^2 /k_{m, \text{min}, i}^2 \big)\big],
\end{align}
with $k_{m, \text{min}}$ the smallest accessible mode in the $m \in \{x, y,z\}$ dimension. For the single-dish experiments, this is determined by the survey's spatial and spectral coverage. For the interferometric HI surveys, we set $k_{x/y, \text{min}}$ to the lower of the smallest mode from the spatial coverage and the primary beam of a single dish (which roughly corresponds to the lower bound on accessible baselines).

The overall window is given as,
\begin{align}
w_i(\bar k) = w_{i, \text{beam}}(\bar k) \; w_{i, \text{hpf}}(\bar k).
\end{align}

The top right panel of Figure \ref{fig:windowmodes} shows these two window functions, along the three independent Cartesian dimensions, for the TIM \sone survey. For the current generation deep-field LIM surveys, there is a distinct mismatch in the modes probed by the line-of-sight and transverse directions, with all $k$-scales affected by either the spatial or spectral window functions. 
Given the theoretical power spectrum $P^t_{ij}$, the surveys measure the windowed version $P^m_{ij} = W_{ij} \times P^t_{ij}$, with $W_{ij} = w^*_i w_j$. 

The uncertainty in the true signal of the cross-power spectrum is given as
\begin{align}
\sigma^t_{ij}(\bar{k})^2 = \frac{1}{2 N_{ij}} \big[& (P^t_{ij})^2 +
(P_{ii}^t + P^n_{ii}/W_{i})(P_{jj}^t + P^n_{jj}/W_{j}) \big],
\label{eqn:noisecross}
\end{align}
with (defined at a given $z$ and $\bar k$) the cross-power theoretical signal $P^t_{ij}$ for two tracers $i, j$, the effective number of modes $N_{ij}$, the auto-power theoretical signals  $P^t_{ii / jj}$, and the noise and window in the auto-power spectra for each survey $P^n_{ii / jj}$ and $W_{i / j}$. We derive this expression in Appendix \ref{appendix:lim}. 

In Section \ref{sec:results}, we express our $\mathrm{S/N}$s in this true (not measured) signal:
\begin{align}
\mathrm{S/N}(k, z) = P^t_{ij} / \sigma^t_{ij},
\end{align}
while the quoted total S/N for an experiment at a redshift is the square root of the quadratic sum over $k$.

\subsection{Computing S/N for deep-field pencil-beam surveys with \texorpdfstring{$k$}{k}-space lattice construction}

To compute the relevant quantities in Eqn. \ref{eqn:noisecross}, we detail a formalism for deep-field LIM surveys, i.e, pencil-beam-like maps made by medium-resolution spectrometers $R\sim10^{2-3}$. These datasets exhibit high anisotropies between the $k$ modes probed by the line-of-sight and transverse directions, due to a mismatch in both resolution and coverage. Typically, the line-of-sight coverage and smallest resolved element are much higher than in the transverse direction (in equivalent units), or equivalently $\kparamin << \kperpmin$ and $\kparamax << \kperpmax$. 
Past works \citep{Furlanetto_2007, Li_2016, Bernal_2019, Padmanabhan_2022, Padmanabhan_2023} write the number of modes $N_\text{modes}(k)\propto k^2 \Delta k \;V_\text{survey}$ and/or perform unweighted integrals over $d\mu = d(\cos\theta) = d(\kpara / |\bar k|)$ over $(0, 1)$. These expressions implicitly assume $k$-space isotropy. Instead of writing analytical expressions over $k = |\bar{k}|$ for intermediate quantities (e.g., noise per mode or window contribution), we work in the three-dimensional $\bar k$ space by explicitly constructing the $k$-space lattice cuboid that is accessible to each of the considered LIM surveys. 

In $k$-space, the smallest resolution element is $k_{m,\text{min}}$ and the largest fundamental mode is $k_{m,\text{max}} = 2\pi / \delta l_m$ ($m \in \{x, y, z\}$). Thus, with $L_m$ and $\delta l_m$ the largest and smallest mode accessible in dimension $m$, the full $k$-space lattice is given as:
\begin{align}
\label{eqn:lattice}
\bar{k} = (&n_x k_{x,\text{min}}, n_y k_{y,\text{min}}, n_z k_{,\text{min}}),
&\text{with $k_{m, \text{min}} = 2\pi / L_m$, \; $\forall \;n_m \in \mathrm{Z}, |n| \leq L_m / \delta l_m $},
\end{align}

This formalism enables averaging anisotropic functions, such as those of the form $f(\bar{k})$ or $f(k,\mu)$, by computing over the lattice distribution; marginalizing over $k$ or $\mu = \cos\theta = \kpara / |\bar k|$ ensures correct weighting given survey parameters and skew. 

For example, the binned mode counter $N_{i}(k)$, number of modes with $\bar k$ magnitude in ($k$, $ k + \delta k$), is:
\begin{align}
N_{i}(k) &= \sum_{|\bar{k}| \in (k,  k + \delta k)} 1,  
\end{align}
that is, the number of lattice points within the thin three-dimensional spherical shell of thickness $\delta k$ at radius $k$. We use this expression for the \halpha galaxy counts survey. For the line intensity mapping surveys considered, we actually use
\begin{align}
N_{i}(k) &= \sum_{|\bar{k}| \in (k,  k + \delta k)} Z(\bar k),  
\end{align}
where $Z(\bar k) = 0$ \textit{iff} $k_x =0$ or $k_y=0$ or $k_z=0$, and 1 everywhere else. This conservatively removes modes with a constant component along a Cartesian axis, which could potentially be contaminated by effects that make extracting the true astrophysical signal intractable \cite{Furlanetto_2007}.  Fiducial examples include \textit{Herschel}'s broadband maps, which are mean zero, i.e., the true DC astrophysical signal (sky-averaged emission) was unrecoverable, as well as interferometric maps.
For calculating sensitivities to cross-power spectra, we conservatively select lower mode counts between the two tracers in two-dimensional bins over $\mu=\cos\theta=k_z/k$ and $k$; we bin modes in $\mu-k$ space (with $\delta \mu = 0.1$ and $\delta \log k = 0.3$) to obtain $N_i(k, \mu)$ and $N_j(k, \mu)$, select $N_{ij}(k, \mu) = \min(N_i, N_j)$, and then sum over $\mu$: $N_{ij}(k) = \sum_\mu N_{ij}(k, \mu)$. The top left panel of Figure \ref{fig:windowmodes}
compares the number of modes at different $|\bar k|$ given by the fiducial analytical expressions (scaling with the volume of thin spherical shells of radius $|\bar k|$) and our $\bar k$-space lattice-based mode counting; it also showcases the effect of conservatively excluding zero-component modes, notably at larger scales for an anisotropic survey volume like that of the TIM \sone survey. 

The bottom panel of Figure \ref{fig:windowmodes} shows the effect of applying our windowing and mode counting formalisms to the noise spectrum for the $z\sim1.17$ redshift bin of the TIM \sone \ciium survey. The windowing suppresses power at all scales, substantially at smaller scales due to the line-of-sight component of the window. Similarly, the lattice-based mode counting reduces the number of modes at high $|\bar k|$ (as the survey only samples a small section of the fiducial spherical shell in $k$-space), suppressing sensitivities in the shot-noise regime.

\subsection{Survey noise in power spectrum measurements}

Expressions for the noise power spectrum depend on the survey type. For a galaxy catalog, it is the systematic shot noise contribution,
\begin{align}
P^n_\text{gal} = 1 / n_\text{gal},
\end{align}
where $n_\text{gal}$ is the number density of emitters detected by the galaxy catalog.

For single-dish LIM surveys with uniform field coverage, given a noise equivalent intensity (NEI) $\sigma_n$ (in Jy/sr$\sqrt{s}$), we can write the per-voxel noise in the real-space map (assuming non-white Gaussian noise, e.g., detector 1/$f$ or thermal stage drift, has been filtered out):
\begin{align}
\sigma_\text{voxel}^2 = \frac{\sigma_n^2}{t_\text{voxel}}; \text{ with, } t_\text{voxel} = t_\text{survey} N_\text{spaxels} \frac{\Omega_\text{beam}}{\Omega_\text{survey}}.
\end{align}
Here, $t_\text{survey}$ is the total survey observation time, $N_\text{spaxels}$ is the number of independent spatial pixels (concurrent beams) on the sky, and $\Omega_\text{beam}$ and  $\Omega_\text{survey}$ are the beam and survey area on sky, respectively. We also write the three-dimensional voxel volume as:
\begin{align}
V_\text{voxel} = \chi(z)^2 \Omega_\text{beam} \times \frac{d\chi}{d\nu}(z) \delta\nu.
\end{align}
Then, the 3D noise spectrum for a single dish experiment is given as:
\begin{align}
\label{eqn:singledish_Pn}
P_n(\bar{k}) = \sigma_\text{voxel}^2 V_\text{voxel} = \frac{\sigma^2_n}{t_\text{survey} N_\text{spaxels}} \chi^2 \frac{d\chi}{d\nu} \Omega_\text{survey} \delta\nu,
\end{align}
which is isotropic and constant in $\bar k$ space.

An equivalent expression can be derived for an interferometric experiment, noting that observing strategy and antenna placement determine the number density of baselines $n(u, z)$, which affects sensitivity to different $\kperp$; see \cite{Bull_2015} for a full derivation. Given the total system temperature $T_\text{sys} = T_\text{sky} + T_\text{inst}$, and assuming the $uv$-space resolution element $\delta u\,\delta v \sim 1/\Omega_\text{FOV}\sim A_e/\lambda^2$, we can write:
\begin{align}
P_n(\bar k) = P(\kperp, \kpara)=  \frac{(2k_B T_\text{sys})^2}{2 \; t_\text{survey} \; A_{e}^2}  \chi^2 \frac{d\chi}{d\nu} \frac{\Omega_\text{survey}}{\Omega_\text{FOV}} \frac{1}{n(u, z)},
\end{align}
in units of $\text{Jy}^2\text{Mpc}^3$, assuming number of independent polarizations $n_\text{pol}=2$, with effective collecting area of a dish $A_e$, field of view $\Omega_\text{FOV}$, and conversion from temperature to intensity in the Rayleigh-Jeans limit $I\sim 2k_B T_\text{sys} / \lambda^2$. In the limit of long observations, the baseline density $n(u, z)$ is taken to be uniform.

A crucial quantity, thus, for writing down the noise power spectrum of an interferometer is the number density of baselines in the visibility $u-v$ plane, $n_b(u, z)$.
Past works have simply approximated the baseline distribution in the $u-v$ plane to be uniform \citep{Bull_2015, Bernal_2019, Padmanabhan_2023}, spread in a circular disk whose radius is set to the maximum baseline separation of $b_\text{max}\sim 8$ km,
essentially spreading $N_\text{total}(N_\text{total}-1)/2$ baselines over an area of $\pi b_\text{max}^2$.
However, this is a poor assumption, particularly in the case of MeerKAT or SKA-Mid with a large core sub-group of antennae that are limited to the inner $\sim 1-2$ km; 
for MeerKAT, this is $N_\text{inner}=44$ of the $N_\text{inner}+ N_\text{outer} = N_\text{total} = 64$ antennae. 
While the exact distribution of baselines in the $u-v$ plane is dependent on a specific survey's scan strategies \footnote{e.g., see \href{https://apps.sarao.ac.za/calculators/uvbeam}{apps.sarao.ac.za/calculators/uvbeam}}, the density of baselines is higher in this inner (smaller-area) disk of radius $\sim 2$ km. 
For this analysis, we considered all the baselines subtended by the inner core of antennae ($N_\text{inner}(N_\text{inner}-1) / 2$) and half the baseline where only one of the antennae was from the core ($N_\text{inner}(N_\text{outer}) / 2$),
equally spread over a smaller disk of radius $b^c_\text{max} = 2$~km (i.e., ignoring far-spaced antennae-pairs subtending $\gtrsim 2$ km baselines). While this is a crude approximation as the $u-v$ density of baselines is anisotropic and heavily dependent on observation strategy, RFI (radio frequency interference) removal strategy, and 
field choice, our approximation is closer to a realistic $u-v$ coverage than a simple uniform density model over the entire available $u-v$ coverage.
For MeerKAT, we thus set $n_b(u, z) \sim 1.32 \times 10^{-4} \times \lambda_\text{obs}^2$ m$^{-2}$, with $\lambda_\text{obs}= 21~\text{cm} \;(1+z)$. For SKA-Mid, we take $N_\text{inner}\sim N_\text{total} / 2$, and get $n_b(u, z) \sim 7.64 \times 10^{-4} \times \lambda_\text{obs}^2$ m$^{-2}$, i.e., roughly a $\sim6\times$ increase in baseline density within the core 2~km radius. The minimum transverse scales accessible are also set by this 2~km maximum baseline.

\section{Results: Forecasts for Nine Power Spectra \label{sec:results}}

We present forecasts on three auto-power spectra and six cross-power spectra signals, in four different redshift bins for the three different generations of experiments described in Section \ref{sec:surveys}. 
The formats of Figures \ref{fig:sensitivitiesauto}, \ref{fig:sensitivitiesgal}, and \ref{fig:sensitivities} are identical. Each row corresponds to one of these nine signals in each of the four redshift bins. Each panel displays the median of the theoretical power spectrum, along with the $16^\text{th}-84^\text{th}$ percentile model space interval, resulting from intrinsic scatter and empirical uncertainty in source luminosities (see Section \ref{subsec:simim}). For each $z$, three curves show the per-$k$-bin $1\sigma$ sensitivity for the three \sonens, \sfourns, and \shund survey configurations. As noted in Section \ref{subsec:meerkat}, MeerKAT UHF receivers are not sensitive to the 21 cm line above $z\sim1.5$; the dotted curves in the highest $z$ bin for these surveys hypothesize a 64-dish configuration fitted with the SKA-Mid Band 1 receivers.

\subsection{Auto-Power Spectra \label{subsec:auto} }

\begin{figure*}[hp]
\centering
\includegraphics[width=1\linewidth]{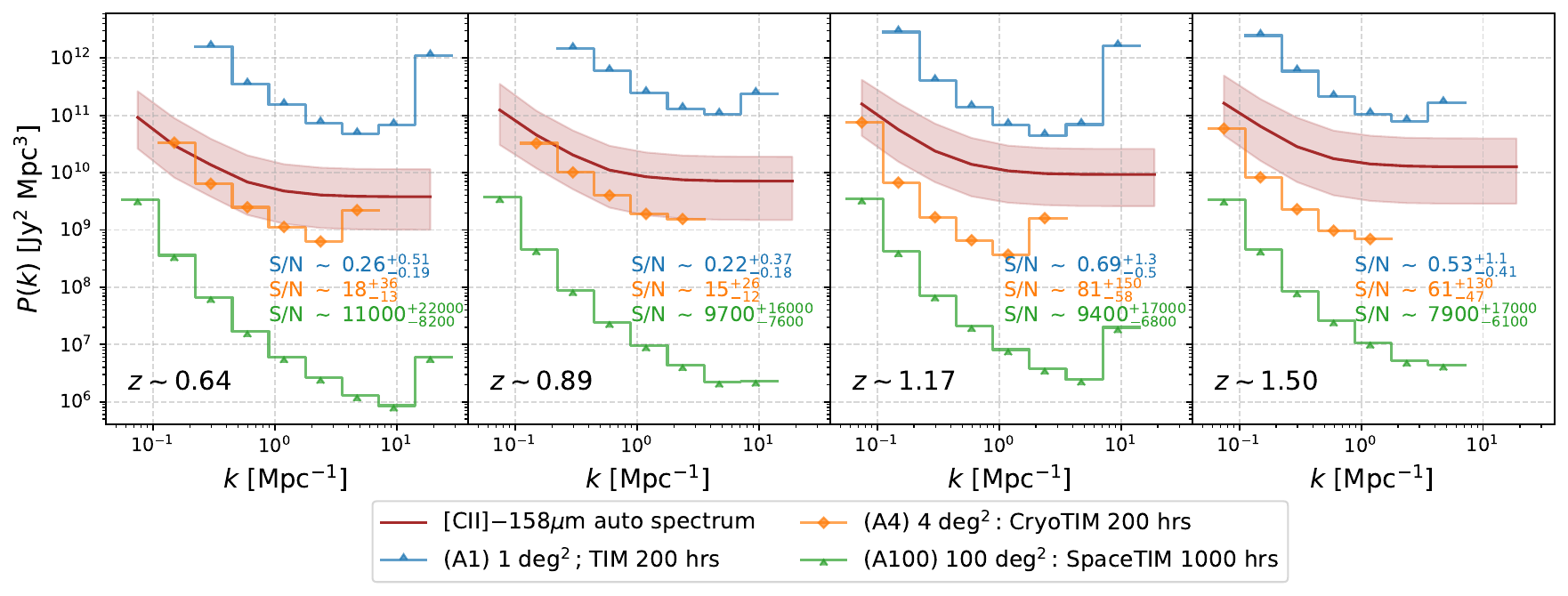}
\includegraphics[width=1\linewidth]{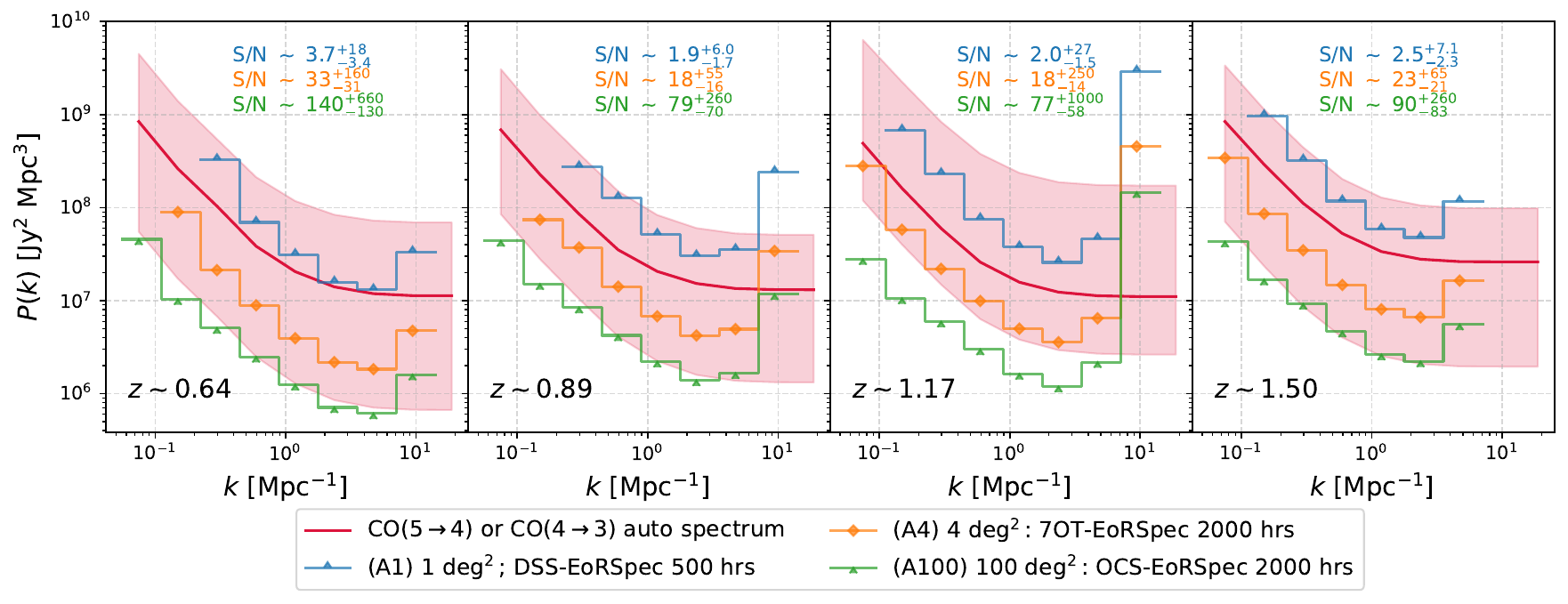}
\includegraphics[width=1\linewidth]{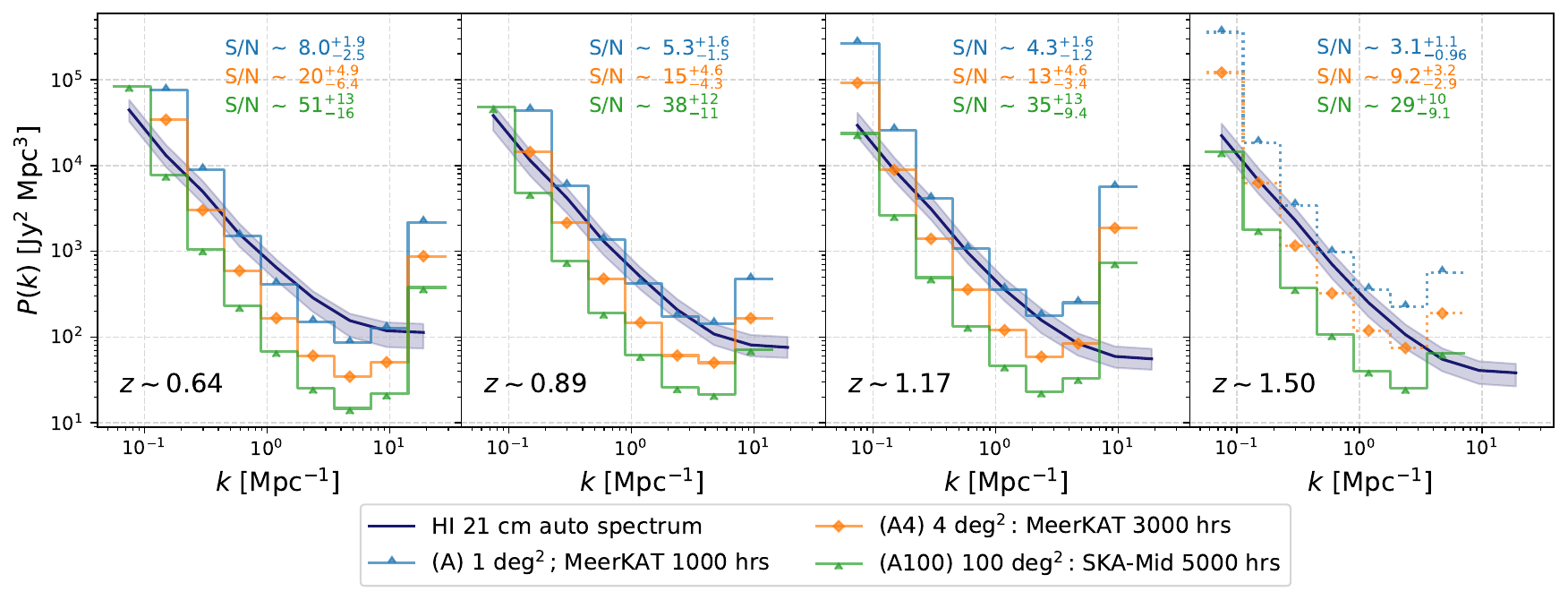}
\caption{\label{fig:sensitivitiesauto}
Forecasts for sensitivities to the line intensity auto-correlations, in four redshift bins over $z\sim0.5-1.7$, for \ciiumns, CO$_{J+1\rightarrow J}$ ($J=3, 4$), and HI 21 cm, for the three classes of surveys considered with coverage $1\deg^2$, $4\deg^2$, and $100\deg^2$. The scatters shown in the theoretical spectra (as well as the uncertainties on the signal-to-noise ratio S/N) are model uncertainties, marking 16th, 50th, and 84th percentiles. The \sone far-infrared spectrometer, TIM, will place upper-limits on the auto-power spectrum, assuming an SFR-tracing \ciium intensity history; note that certain other models for \cii emission can predict up to two orders of magnitude brighter auto-correlation signal, allowing the current generation TIM survey to discriminate between these models. MeerKAT will be able to place integrated S/N $\sim10$ detections on the HI signal with the as-proposed LADUMA survey, with SKA-Mid approaching higher S/N in individual $k$ bins. 
Even with the $1\deg^2$ portion of the DSS, EoRSpec will place S/N $\sim3$ limits on the CO auto-spectrum, with each of the next two generations offering a $10-100\times$ improvement. 
}
\end{figure*}

Our forecasts for detecting the auto-correlation signals for the three LIM tracers are shown in Figure \ref{fig:sensitivitiesauto}. These additionally motivate the independent constraining power of each of our survey configurations, which is useful for interpreting the cross-correlation forecasts.

We forecast that the current generation of \ciium experiments will resolve the tension between competing models and measurements of the intensity history at $z\sim 1-2$, which vary by over an order of magnitude at cosmic noon. TIM will exclude brighter models, placing upper limits on the auto-correlation signal, assuming \ciium linearly traces star formation rate as suggested by \cite{delooze+14, agrawal2025farinfraredlineshiddenarchival}. A \sfour survey with a background-limited balloon telescope would enable an auto-correlation detection of \ciium with S/N $\sim 20-80$ in 200 hours. A space observatory will push this to S/N $\sim 10^4$ in each redshift bin, with per-$k$-bin sensitivities of S/N $\sim 10^3$, enabling sub-percent level constraints on the clustering and shot regime power spectra.

We forecast S/N $\sim3-10$ and $\sim 10-20$ for either single-pointing HI survey with MeerKAT. The as-proposed LADUMA observations can, thus, detect the 21 cm auto-spectrum across cosmic noon, and place S/N $\sim 3$ limits on the signal at various scales. 
Additional baseline density in SKA-Mid allows the 100 $\deg^2$ survey to achieve S/N $\sim 50$ and per--$k$-bin S/N $\sim 3-10$.
We forecast the $1\deg^2$ portion of the DSS with EoRSpec to detect the CO auto-spectrum at S/N $\sim2-3$; note that the full $2\deg\times2\deg$ DSS survey will be up to twice as sensitive, and also measure larger spatial modes on the sky.  The \sfour and \shund surveys offer over an order of magnitude improvement in S/N, correlating with the corresponding increase in spectrometer-hours on sky.
For all three tracers, we see the accessible $k$-modes evolve with $z$; this effect is most notable as field sizes increase with survey class, and for CryoTIM with a much smaller primary mirror than its analogues. 

\subsection{Cross-Power Spectra: Line$-$Galaxy and Line$-$Line}

\begin{figure*}[hp]
\centering
\includegraphics[width=1\linewidth]{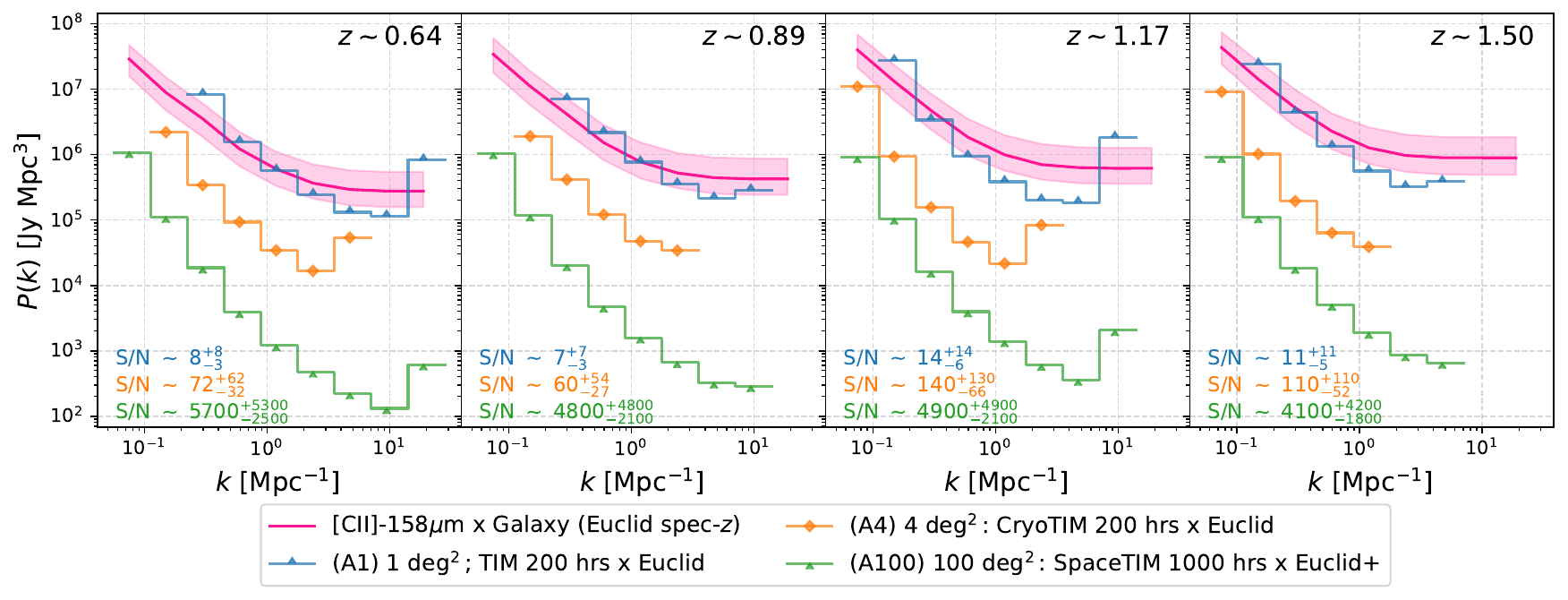}
\includegraphics[width=1\linewidth]{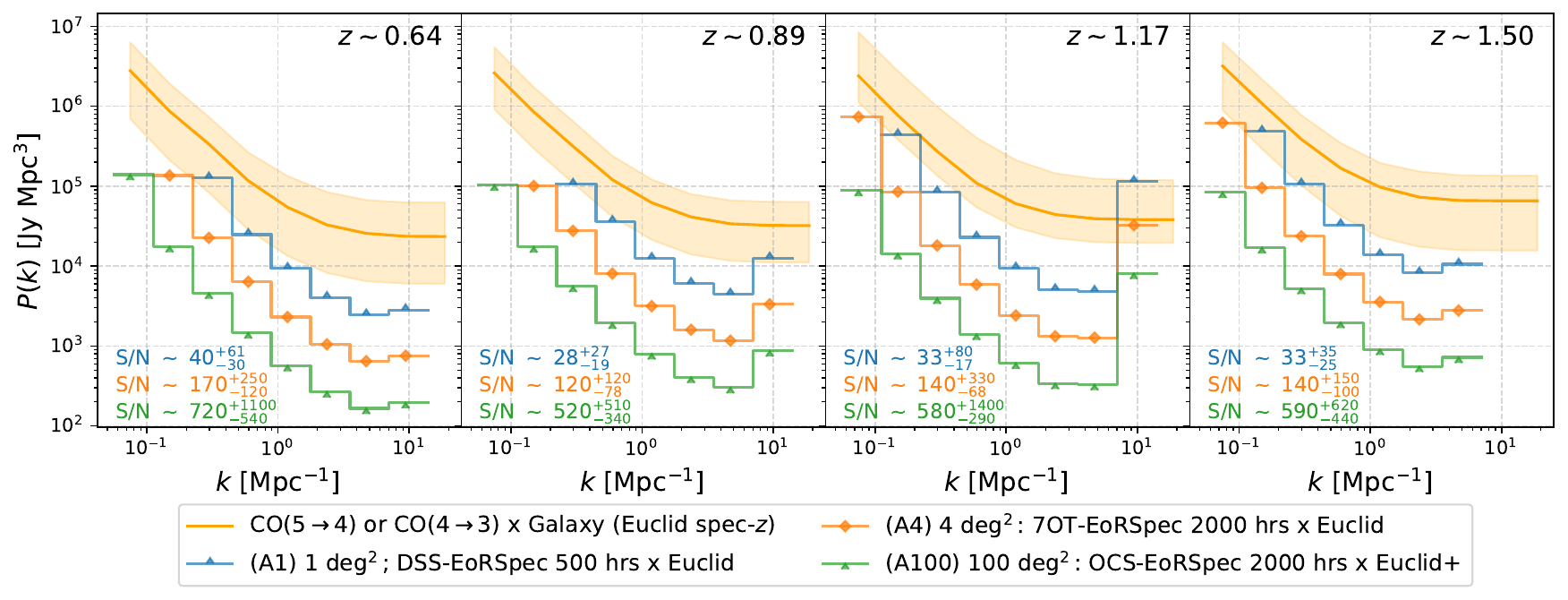}
\includegraphics[width=1\linewidth]{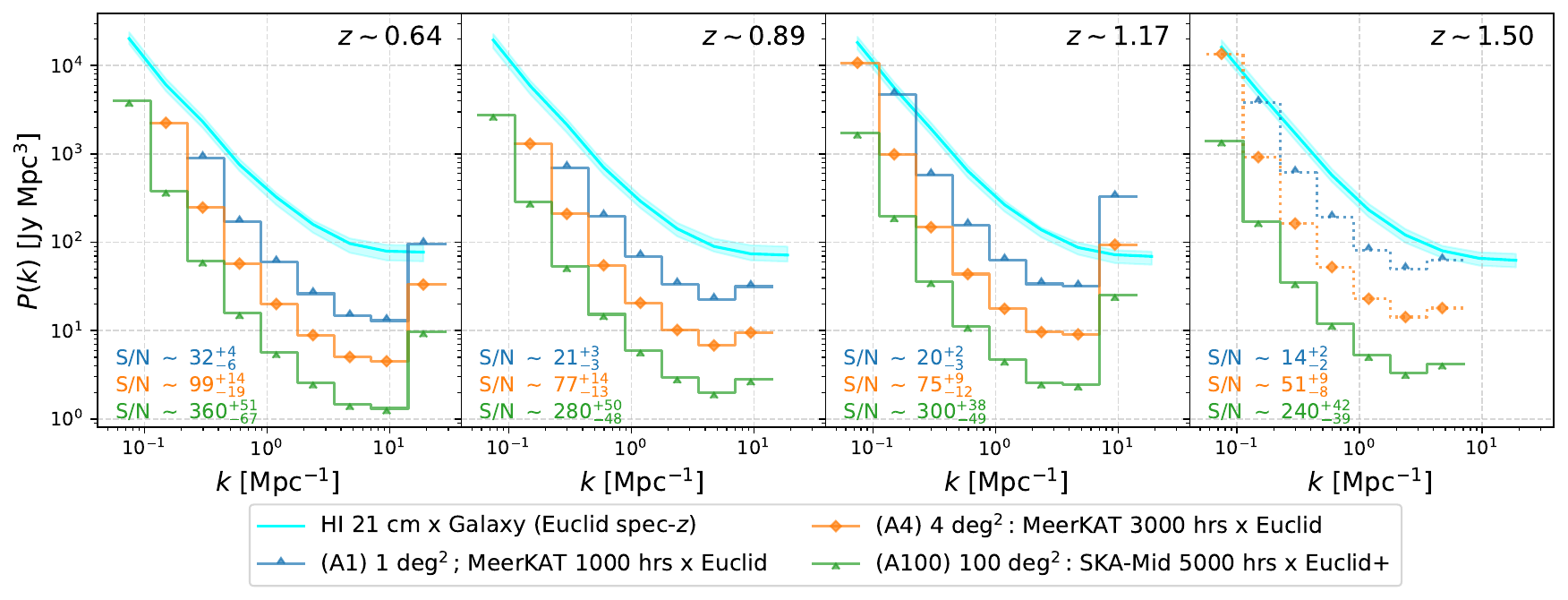}
\caption{
\label{fig:sensitivitiesgal}
Our forecasts for sensitivities to the LIM \ciiumns, CO, and HI 21 cm cross-correlations with \textit{Euclid}-like spectroscopic galaxy surveys, over $z\sim0.5-1.7$, for the three classes of surveys considered with coverage $1\deg^2$, $4\deg^2$, and $100\deg^2$. The galaxy density fields are constructed from boolean point clouds of sources detectable given a \halpha deep-field flux limit for \textit{Euclid}, and offer a high S/N tracer for extracting line intensity signal in cross-correlation. All three LIM tracers have their signals amplified by $\sim3-30\times$. This is exceptionally significant for current-generation surveys, making the integrated signal detectable for \ciium at S/N $\sim 10$ and the per-$k$-bin signal detectable for HI 21 cm and CO at S/N $\sim 3$.
}
\end{figure*}

Our forecasts for cross-correlation of LIM observables with a \textit{Euclid}-like spectroscopic galaxy survey are shown in Figure \ref{fig:sensitivitiesgal}. While the galaxy surveys are dependent on observer-frame \halpha line intensities, they are \textit{not} a line intensity mapping survey per se, as the field is number counts (after a flux cut) and not intensities. Thus, information about the distribution of \halpha intensities is lost, and emission is only accounted for from identified sources. 

The most significant S/N boosts from galaxy cross-correlation 
are for the \sone and \sfour \ciium surveys. The cross-power spectrum for \sone (TIM $\times$ \textit{Euclid}, $1\deg^2$) is potentially detectable with S/N $\sim 10-20$, particularly at the higher $k$-modes in the shot noise regime. The per-$k$-bin S/N is still only $\sim1-3$; the background-limited \ciium \sfour survey could yield S/N $\gtrsim10$ in several individual $k$-bins, mapping both the clustering and shot-dominated power spectrum robustly. The \shund survey yields another improvement of over an order of magnitude, again offering sub-percent cosmological measurements like the auto-correlation. 
CO and HI are both $0.5 - 1.5$ dex ($\sim3-30\times$) more detectable in cross-correlation with galaxy surveys, pushing to S/N $\sim30$ with the $1\deg^2$ fields. The per-$k$-bin sensitivities are similarly boosted, enabling S/N $\sim 3-10$ measurements at various angular scales with current generation surveys. Deploying each next class of instrumentation for 21 cm and CO LIM enables a $0.5-1$ dex ($\sim 3-10\times$) improvement in sensitivities to the galaxy cross-correlation, a combined consequence of the increased depth and field size. 

Our power spectrum model and S/N forecasts for \ciium $\times$ \halphans-galaxy counts for a TIM and \textit{Euclid} cross-correlation are consistent with complementary work \citep{timconstraints}, which does not use an MHD-simulation-based forward-modeling methodology. This serves as an additional independent validation for our calculation.
\cite{timconstraints} constructs their cross-spectra model in Fourier space by analytically decomposing the signal into the two-halo clustering and shot-noise terms (as per Eqn. \ref{eqn:analytic} and the corresponding discussion), using estimates for the linear bias and shot contributions and offering a faster, lightweight implementation.  
We model the signal in real space, making fewer simplifying assumptions, and include the one-halo contribution; the \cite{timconstraints} model deviates from the cross-spectrum at small scales ($k\sim1-10$ Mpc$^{-1}$) with the inclusion of the sub-structure within halos, but their predicted signal is still within our 68 percentile interval. 

In the cross-correlation setup, wider fields yield higher S/N for all three tracers. This is not a given for auto-correlations, where a wider field for the same instantaneous sensitivity implies a penalty on the instrument-noise spectrum, when cosmic variance does not dominate the noise budget. Crudely, $N_\text{modes}$ scales quadratically with survey spatial (on-sky) solid angle $\Omega$, while $P_{ii}^n\propto 1/\Omega$. So, in the limit that $P_{ii}^n >> P_{tt}^t$, survey on-sky coverage does not affect the S/N of a cross-correlation. This is only true at $k$ modes with no significant windowing; in this specific deep-field setup, an increased field size results in more favorable windowing and cross-linking between accessible line-of-sight and transverse modes, typically resulting in an improvement in S/N by a factor of $\lesssim5$.

\begin{figure*}[hp]
\centering
\includegraphics[width=1\linewidth]{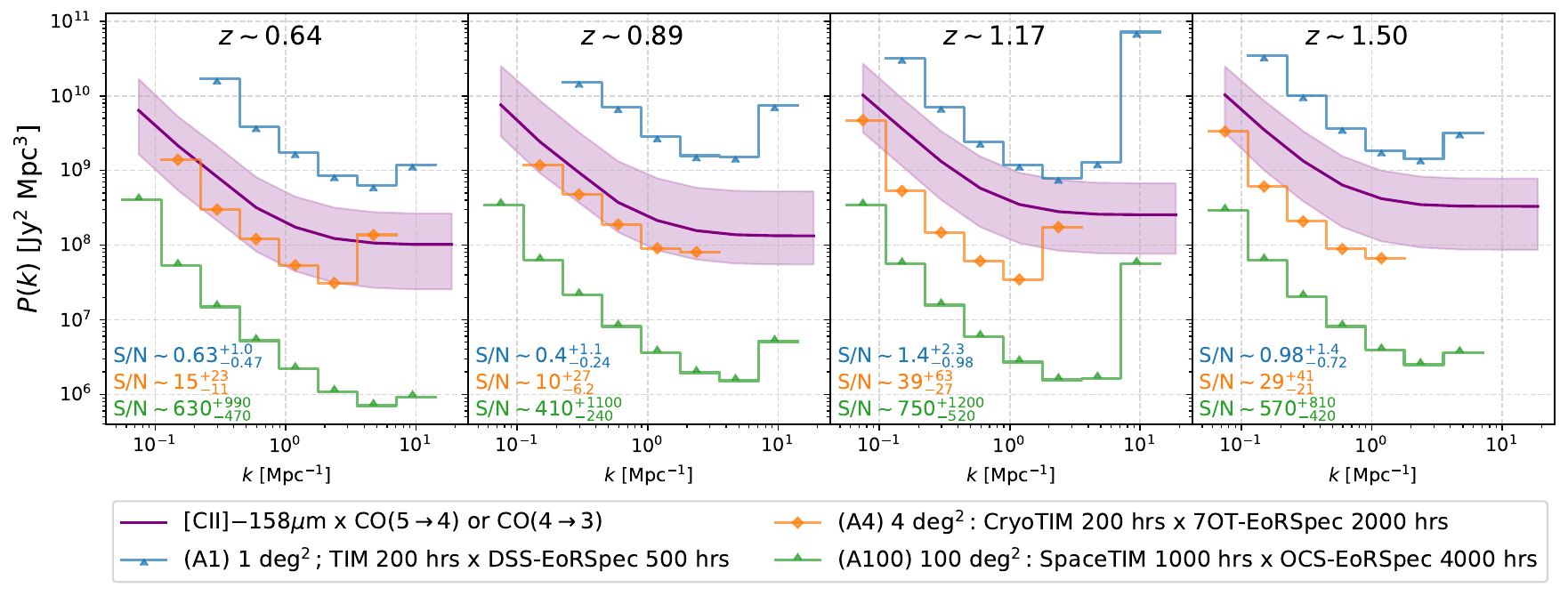}
\includegraphics[width=1\linewidth]{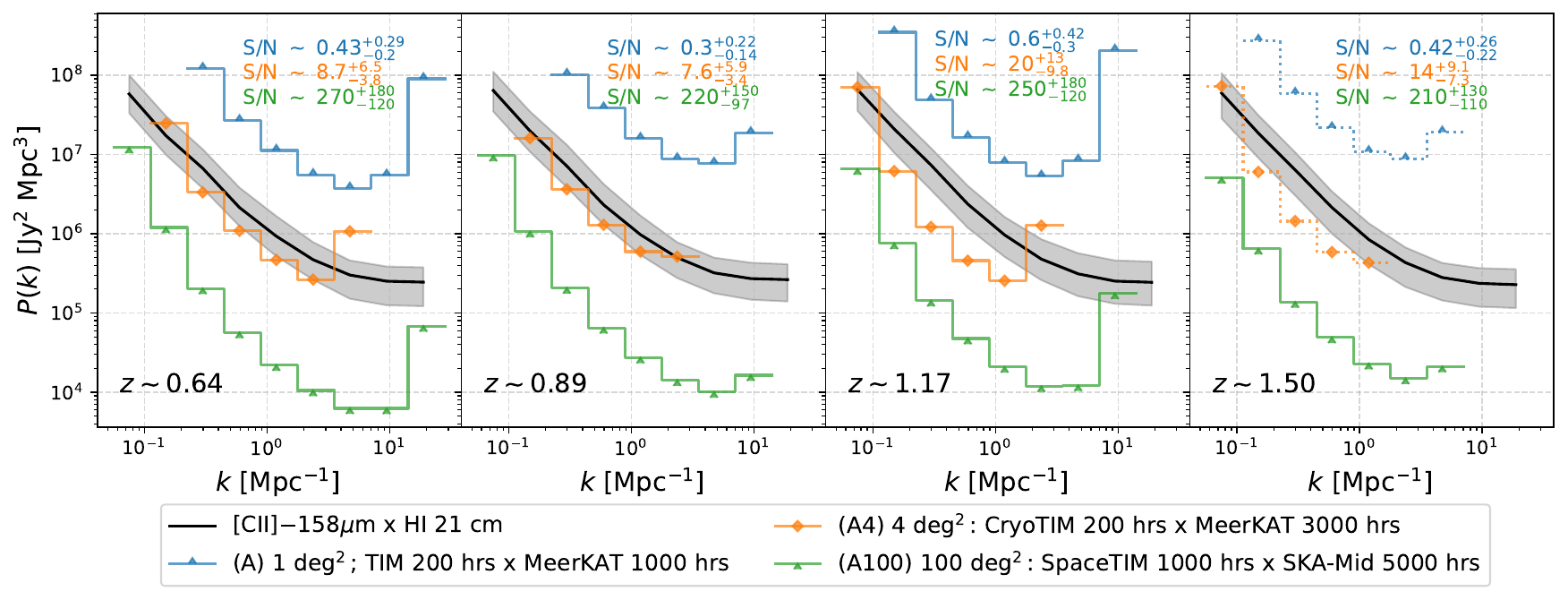}
\includegraphics[width=1\linewidth]{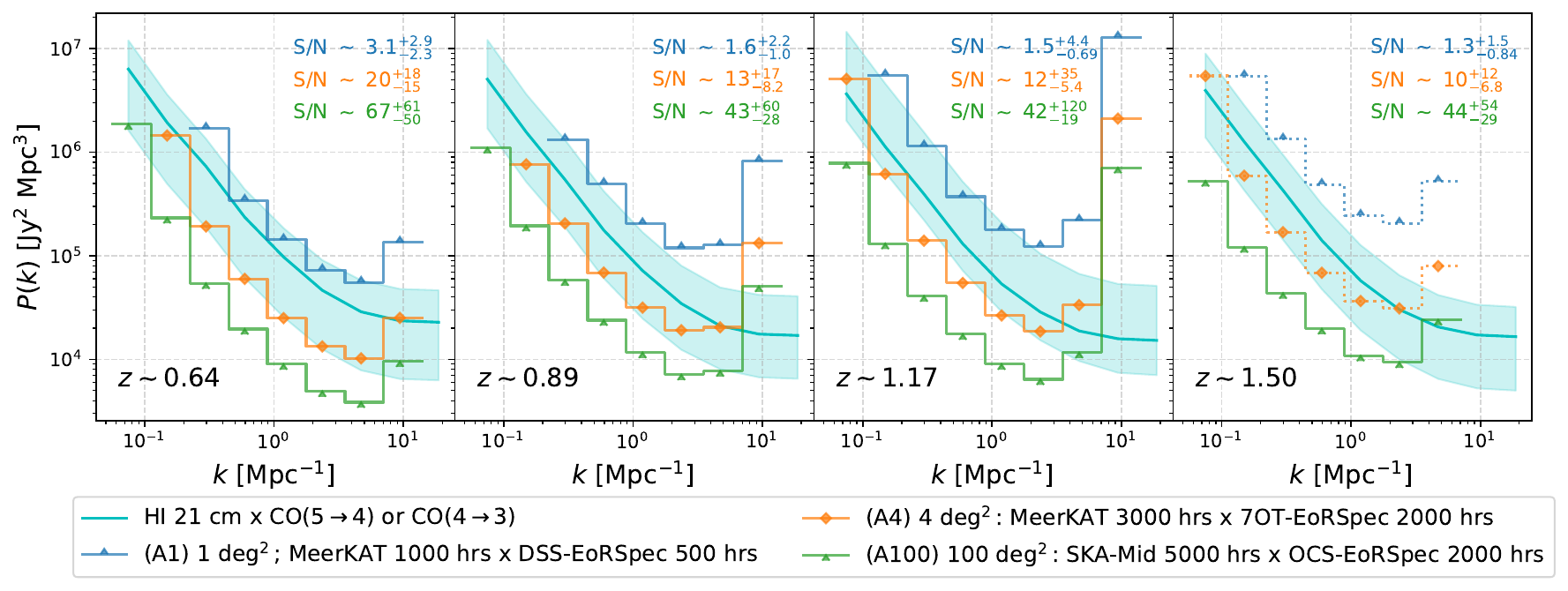}
\caption{Our forecasts for sensitivities to the multi-line cross-correlations between \ciiumns, HI 21 cm, and CO$_{J+1\rightarrow J}$ ($J=3, 4$). Cross-correlations with \ciium are not forecasted to be detectable with the current generation of surveys; but, the upper limits placed will already be useful due to uncertainty in models for \ciium and the relevant measurement tensions at $z\sim1-3$. A suborbital background-limited spectrometer is necessary to detect these signals, assuming the lower intensity \ciium models. 
The CO $\times$ HI 21 cm spectrum can be detected by the current generation of instruments at S/N $\sim3$, with reasonable updates enabling detection at various $k$ scales.
}
\label{fig:sensitivities}
\end{figure*}

This is also evident when considering the cross-correlation of two LIM survey datasets, as shown in Figure \ref{fig:sensitivities}. 
As the \ciium density field measurements with the current generation are at a low S/N in auto-correlation (see discussion in Section \ref{subsec:auto}), their cross-correlation with the other two LIM tracers is not forecasted to be detectable by the \sone class of surveys. However, similar to the auto-spectra, the upper limits on these statistics from TIM will exclude the order-of-magnitude brighter models of \ciium intensity at these redshifts.
A background-limited suborbital survey would enable the detection of these line-line cross-spectra at S/N $ \sim 10$. Combining a space platform's survey with those of futuristic millimeter-wave and radio observatories will enable overall S/N $ \sim 200-800$ and S/N $\gtrsim 5 - 10$ in individual $k$-bins. 
The CO $\times$ HI 21 cm is detectable at S/N $\sim3$ by the \sone surveys, with the next-generation instruments offering S/N $\sim 50-100$ and per-$k$-bin S/N $ \gtrsim 5$.

{
\begin{sidewaystable}
\centering
\renewcommand{\arraystretch}{2}
\centering
\begin{tabular}{c||r|r|r|r}
Statistic & \sone S/N & \sfour S/N & \shund S/N & Astrophysics Informed
\\
 & (Current) & (Next-Gen) & (Futuristic) & (... + LSS + line-line covariance + $P_\text{shot}$)
\\
\hline
\cii auto
& $0.69^{+1.3}_{-0.5}$ & $81^{+150}_{-58}$ & $9400^{+17000}_{-6800}$
& cosmic SFRD, SF clustering, $\langle I_\text{\ciins} \rangle$, $b_\text{\ciins}$, sSFR, SFR MS, SFRF
\\
\cii$\times$ Galaxy
& $14^{+14}_{-6}$ & $140^{+130}_{-06}$ & $4900^{+4900}_{-2100}$
& C$^+$–halo occupation, $L_\text{\ciins}(M_h)$, SFR/$M^*$-\ciins, duty cycle
\\
\hline
CO auto 
& $2.0^{+27}_{-1.5}$ & $18^{+250}_{-14}$ & $77^{+1000}_{-58}$
& molecular gas abundance $\Omega_\text{H$_2$}$, CO excitation $r_{J, J'}$, $b_\text{CO}$, $\alpha_\text{CO}$, H$_2$ MF
\\
CO $\times$ Galaxy
& $33^{+80}_{-17}$ & $140^{+330}_{-68}$ & $580^{+1400}_{-290}$
& H$_2$$-$halo scaling, $M_\text{H$_2$}(M_h)$, $L_\text{CO}(M_h)$, molecular gas cross-bias
\\
\hline
HI auto
& $4.5^{+1.6}_{-1.2}$ & $15^{+5.6}_{-4.1}$ & $53^{+19}_{-14}$
& diffuse cold gas reservoir, HI MF, $\Omega_\text{HI}$, $b_\text{HI}$, $\langle I_\text{21cm} \rangle$, cosmic refueling
\\
HI $\times$ Galaxy 
& $20^{+2}_{-3}$ & $81^{+10}_{-13}$ & $350^{+46}_{-57}$
& HI halo occupation, $M_\text{HI}(M_h)$, $L_\text{HI}(M_h)$, SFR/$M^*$-HI, cross-bias
\\
\hline
\cii$\times$ CO
& $1.4^{+2.3}_{-0.98}$ & $39^{+63}_{-27}$ & $750^{+1200}_{-520}$
& SF efficiency, $\tau_\text{depletion}$, dense gas–SFR covariance, SFR-scaling
\\
\cii$\times$ HI
& $0.62^{+0.43}_{-0.3}$ & $23^{+15}_{-11}$ & $300^{+210}_{-150}$
& gas conversion efficiency, cross-bias, accretion$–$SF coupling
\\
HI $\times$ CO
& $1.5^{+4.5}_{-0.7}$ & $13^{+39}_{-6.1}$ & $58^{+170}_{-26}$
& HI$-$H$_2$ joint occupation, phase balance, H conversion efficiency
\end{tabular}
\caption{\label{tab:snrs} An overview of the nine statistics for which we model signals from empirical measurements and forecast sensitivities for three generations of LIM. We list the expected total S/Ns in the redshift bin at $z\sim 1.17$ (co-added over $10^{-1}~\text{Mpc}^{-1} \lesssim |\bar k| \lesssim 10^{1}~\text{Mpc}^{-1}$), along with a compilation of astrophysics constrained by the corresponding statistic. 
All the species trace the underlying dark matter power spectra, providing cosmological information, especially in the clustering regime. The shot noise terms, bias factors, and mean intensities further provide a census of galactic chemistry, star formation, molecular gas, atomic reservoirs, phase conversion, and the baryon cycle --- as well as their redshift evolutions and trends with galaxy properties. We forecast that galaxy cross-correlations are the main discovery channel for current-generation surveys, with next-generations and futuristic surveys enabling percent and sub-percent level constraints on the line auto and cross spectra.}
\end{sidewaystable}
}

\section{Discussion and Astrophysical Potential for Three Generations of LIM \label{sec:discussion}}\

Table \ref{tab:snrs} lists the nine power spectra we consider in this work, along with forecasted total S/Ns at $z\sim 1.17$ for the three generations of LIM surveys. We also list astrophysical quantities informed by measurements of these statistics, in addition to the cosmological constraints from measuring fluctuations in the clustering regime. \ciiumns, CO, and HI 21 cm trace SFR, molecular gas, and atomic hydrogen, providing a census for components of the baryon cycle and their evolution with redshift. The cross-power spectra constraint the correlation strengths of each pair of tracers, encoding trends with galaxy properties and line-line covariance; these inform models of star formation efficiency, halo occupation, gas temperature and depth, and the coupling of cosmic accretion to the baryon cycle. 

Cross-correlations with spectroscopic galaxy survey counts will enable current generation experiments to place $\mathrm{O}(10)$ S/N constraints on \ciiumns, CO, and HI line intensity histories at $z\sim 0.5-1.7$, improving on auto-correlation S/N by factors of $\sim3-30\times$. While a Fischer analysis for forecasting constraints on astrophysical observables (e.g., SFR,  total molecular gas density, SF efficiency, depletion time, SF surface density) is left for future work, qualitatively, this will constrain models of the intensity history of these three LIM tracers, inform correlations between galaxy properties (e.g., SFR, halo mass, stellar mass) and species abundances (constraining the model space considered), and aid in the development of priorities and designs for future surveys. We forecast that the intermediate-generation experiments will provide percent-level constraints on cross-correlations, with the next-generation surveys providing the same for auto-correlations.

In addition to a potential S/N improvement, the cross-power spectra encode information complementary to or absent from auto-power spectra. In the two-halo clustering regime, the two auto-spectra independently constrain the two products of the linear bias and mean intensity, $b_1 \bar I_1$ and $b_2 \bar I_2$, while the best constrained quantity by the cross-spectrum is the anti-correlated product $b_1 b_2 \bar I_1  \bar I_2$. This results in different best-constrained directions in the $b_1\bar I_1 - b_2\bar I_2$ space; thus, combining posteriors from all three can yield improved constraints. These linear biases inform models of halo occupation and characteristic source properties. The mean intensities provide a cosmic census of star formation history, molecular gas density, and neutral atomic hydrogen density, informing redshift evolution of the baryon cycle. 
Both are also necessary to constrain to enable next-generation cosmological measurements with LIM tracers. 
In the shot Poisson-dominated regime, the cross-correlation coefficient (see Figure \ref{fig:ccc}) encodes the second moments of the line LF \citep{Schaan_2021}, which, in turn, for our empirical models, encode the differences in scaling factors, in addition to astrophysical intrinsic scatter and line-line covariance. Thus, a combined measurement of two auto-power and the cross-spectra can constrain scaling relations between galaxy properties (e.g., SFR, stellar mass, halo mass) and line luminosities, informing models of galaxy evolution by providing a  census of molecular and atomic gas and their trends with galaxy properties. Notably, even though a galaxy survey like \textit{Euclid} misses low-SFR populations, the clustering signal (at large scales) still encodes the complete census of a species, in the form of the full integrated LF or mean intensity. The shot regime signal is affected by the depth of a galaxy survey, but could be used to quantify mean emission and, thus, molecular and atomic gas or SFR as a function of source properties.

Power spectra are not the only relevant statistic that enable the extraction of information from LIM datasets. Stacking \citep{Delhaize_2013, roy2024measurementco32lineemission, sinigaglia2024mighteehihigalaxyproperties,agrawal2025farinfraredlineshiddenarchival} is a complementary methodology that amplifies signals at the scale of the instrument's three-dimensional beam, typically with an auxiliary source catalog. This catalog can be a spectroscopic redshift survey or a point source extraction from another LIM survey. Directly cross-correlating two fields does enable skipping any source extraction steps and associated systematics from using a noise floor for source selection. 
Voxel Intensity Distribution (VID) and Conditional VID statistics have been explored as a potential statistic that can capture non-Gaussian information, arising due to gravitational interactions at small scales \citep{Breysse_2017, sabla2024unlockingnewphysicsjoint}. 
Forecasting these statistics is possible within our forward modeling methodology, by constructing volumes or light cones at different redshifts from these MHD simulations that directly encode non-linear gravitational collapse.

Slight improvements to our forecasting calculations are possible and left for future work. Notably, we assume the spectral and spatial beams of each instrument to be Gaussian in Eqn. \ref{eqn:beamwindow}. Deviations from this assumption can change the applied attenuation to modes with high $\kpara$ or $\kperp$. For example, \cite{Marcuzzo_2025} considers the spectral beam for EoRSpec to be Lorentzian, instead of Gaussian, and notes a higher attenuation for high $\kpara$. On modeling the signal, star formation burstiness at $z\sim 1$ \cite{2023ApJ...952..133M, 2025ApJ...994...14P}, especially in low-mass galaxies, can affect the line luminosities assigned to subhalos and thus the shape of the power spectrum, particularly at smaller scales. 
Finally, we assume a spectrally constant instrumental noise within each band of the far-infrared, millimeter-wave, and radio detectors. However, these system temperatures and NEIs have narrow spectral features and continuum modulation, which increase the effective noise in measurements at certain redshift slices (for instance, in thin regions of the EoRSpec LF band with poor atmospheric transmission).

The three classes of surveys we have constructed offer a potential roadmap for line intensity mapping over the next few generations of instrumentation, as a staged proposal to increase instantaneous sensitivity and sky coverage. 
For the considered \cii experiments, while the number of concurrent beams on sky is held constant, the background is reduced by $\sim15\times$ and $\sim50\times$ from \sone to \sfour and from \sfour to \shund, complemented with the increase in sky coverage of $4\times$ and $25\times$. Deploying a far-infrared instrument limited by the atmospheric background at balloon altitudes will result in instantaneous noise that is several orders of magnitude lower than the atmosphere from the ground and more than an order of magnitude better than TIM. Experiments have attempted or proposed to cool down a primary mirror to cryogenic temperatures \cite{Essinger_Hileman_2020, 2021JATIS...7d4004S, Kogut_2020, Kogut_2021, Dahal_2024}, but remain limited to shorter survey times, $\mathrm{O}(10)$ hours, due to much higher rates of cryogen use (crudely, resulting in a $\sqrt{200/10} \sim 4 \times$ reduction in noise depth compared to our CryoTIM). 
In the millimeter-wave and radio, ground-based observatories are limited by photon noise or system temperature, resulting in the path forward to be boosting the number of concurrent spectrometers or antennae on the sky. 
Our \sfour and \shund CO surveys increase the number of simultaneous spectrometer beams by $3.5\times$ and $15\times$, while SKA-Mid increases the number of baselines by $7\times$ over MeerKAT. 
Within each class of instrument, next-generation detectors and readout are under development, 
with a focus on scalability through more compact form factors, decreased power consumption, and better cost per pixel
\cite{2016SPIE.9906E..60T, Endo_2019, Karkare_2020, carpenter2020almadevelopmentprogramroadmap, Wandui_2020, Agrawal_2021,sinclair2022ccatprimerfsocbasedreadout, 2022JATIS...8a1021S, Rouble_2022, Yu_2023, benson2025lownoisefouriertransformspectroscopy, benson2025spectralcharacterizationperformancesptslim}.

While we do not directly consider a survey of \halpha line intensities (instead forecasting for a source-counting survey, which maps a three-dimensional field of galaxy overdensities), SPHEREx \cite{2014arXiv1412.4872D} is expected to detect the \halpha power spectrum up to $z\lesssim3$ \cite{Cheng_2024} with its observations of two deep fields at the ecliptic poles. We do not include SPHEREx in this work, as its deep fields do not overlap with the other experiments discussed. A potential SPHEREx successor, the Cosmic Dawn Intensity Mapper (CDIM) is proposed to have a $\sim30$ $\deg^2$ field in the ECDF-S \cite{cooray2016cosmicdawnintensitymapper} at superior noise performance than SPHEREx, making cross-correlations with \halpha LIM a possibility. 
Similarly, the far-infrared enhanced survey spectrometer (FIRESS) on PRIMA \cite{glenn2025prima, 2025prim.book.....M, 2025JATIS..11c1627B, cheng2025featureintensitymappingpolycyclic} provides an $R\sim100$ analogue of the far-infrared space platform considered in this work, operating at shorter wavelengths ($24$–$235~\mu$m) where the zodiacal background is comparatively higher. In this regime, LIM during cosmic noon can target PAH features and key fine-structure lines from [OI], [OIII], [NII], and [SiII].

\section{Conclusion \label{sec:conclusion}}

We have presented a magnetohydrodynamical simulation-based forward modeling and forecasting methodology to predict LIM cross-correlations between \ciins, CO, HI, and \halpha galaxy counts surveys, considering scatter due to intrinsic luminosity variability and model uncertainty. This directly encapsulates predictions for the two- and one-halo clustering and the shot-dominated Poisson regimes, skipping modeling intermediate components of the three-dimensional power spectra. 
We combine this with a formalism for calculating noise power spectra for pencil-beam-like, highly anisotropic, deep-field surveys, using a $k$-space lattice-based approach. We also detail accounting for systematic windowing arising from low $k$-mode contamination, beam attenuation, and finite survey coverage. Our framework is open-source, flexible, and can be used to model signals from other lines or redshifts (given corresponding sub-grid models) and to forecast sensitivities for other surveys. 

We discuss the potential for detecting the auto- and cross-power spectra for these four tracers across cosmic noon from specific multi-wavelength surveys, spanning ground-based, sub-orbital, and space missions. We have presented sensitivity curves for three classes of experiments to nine different statistics, at several angular scales from $k\sim 10^{-1} - 10^1$ Mpc$^{-1}$, in four redshift bins at $z\sim0.5-1.7$, cataloging a wealth of forecasts for line intensity mapping across the peak of star formation history for the next few decades. 

We find that the current generation of LIM surveys will detect the CO and 21 cm auto-spectra, while informing models of \ciium emission via upper limits on the auto-spectra, heralding line intensity mapping as a promising measurement methodology.
We demonstrate that the \textit{Euclid}-like spectroscopic galaxy surveys from space offer powerful density field measurements to cross-correlate with all three LIM tracers, offering a $0.5-1.5$ dex ($3-30\times$) improvement in sensitivities. The three galaxy cross-correlations and the CO $\times$ 21 cm spectra are detectable with the current generation of instruments, offering new astrophysical information in the form of constraints on cross-shot parameters. 

Future iterations of these surveys can constrain these power spectra up to the percent and sub-percent level, at several redshifts and angular scales, making them powerful precision probes of cosmology and galactic astrophysics. 
Our hypothetical experimental setups offer a staged roadmap in each class of instrumentation. The considered improvements range from instruments with better instantaneous noise performance per beam, to reasonable pathways to exponentially increasing the number of concurrent detectors on the sky. Notably, while a space-platform for far-infrared spectroscopy can enable sub-percent precision cosmology and astrophysics with \ciium LIM, there are still significant gains with potentially cheaper and faster sub-orbital platforms beyond the current generation. 
These forecasting results highlight the need for a coordinated effort in the community to map the same sky footprints and geometries at their highest depths, as well as public availability of relevant line intensity data products. 

\acknowledgments
We thank Jonathan Clarke for insightful comments, as well as the TIM and CCAT collaborations for valuable discussion. 

S.A. and J.A. note that this publication was made possible through the support of Grant 63040 from the John Templeton Foundation. The opinions expressed in this publication are those of the authors and do not necessarily reflect the views of the John Templeton Foundation. S.A.'s work was also supported by the Quad Fellowship.

Work on the Terahertz Intensity Mapper is supported by the National Aeronautics and Space Administration under grants 80NSSC19K1242 and 80NSSC24K1881 issued through the Science Mission Directorate.
Part of this research was carried out at the Jet Propulsion Laboratory, California Institute of Technology, under a contract with the National Aeronautics and Space Administration (80NM0018D0004).

\paragraph{Software Used:} \texttt{astropy} \citep{2013A&A...558A..33A,2018AJ....156..123A}, \texttt{yt} \citep{2011ApJS..192....9T}, \texttt{SimIM} \footnote{\href{https://github.com/shubhagrawal30/simim\_public}{github.com/shubhagrawal30/simim\_public}
} 
\citep{keenan+20simim, Keenan_2022}, \texttt{crossLIM} \footnote{\href{https://github.com/shubhagrawal30/crossLIM}{github.com/shubhagrawal30/crossLIM}} (This Work)

\appendix

\section{LIM Surveys with Highly Anisotropic Geometries and Resolution \label{appendix:lim}}

\subsection{Measurement Bias and Variance in Auto- and Cross-Spectrum \label{appendix:variance}}
We exactly write our formalism for two line intensity mapping surveys $i, j$, which measure the true real-valued scalar fields $\delta_{i, j}^t$ in real space with some (assumed to be Gaussian white) noise, or equivalently their Fourier transforms:
\begin{align}
\label{eqn:delxform}
\delta_{i, j}^m(\bar{x}) &= \delta_{i, j}^t(\bar{x}) + n_{i,j}(\bar{x})\\
\label{eqn:delkform}
\tilde{\delta}_{i, j}^m(\bar{k}) &= \tilde\delta_{i, j}^t(\bar{k}) + \tilde{n}_{i,j}(\bar{k})
\end{align}

Without loss of generalization for independent measurements ($i\neq j$), we assume that the noise vectors are uncorrelated, i.e., $\langle \tilde{n}_{i} \tilde{n}_{j} \rangle = 0$. Note that the signal vectors can trace the same or different fields, and are assumed to be uncorrelated with the noise vectors. Thus, the \textit{measured} auto- and cross-power spectra are:

\begin{minipage}{0.45\textwidth}
\begin{align}
\nonumber
P_{ii}&^m(\bar{k}) = \langle\tilde\delta_{i}^{m *} \tilde\delta_{i}^m\rangle = \langle(\tilde\delta_i^t + \tilde{n}_i)^* (\tilde\delta_i^t + \tilde{n}_i)\rangle \\
\nonumber
&= \langle \tilde\delta_i^{t *} \tilde\delta_i^{t} \rangle 
+ \langle \tilde\delta_i^{t *} \tilde n_i \rangle 
+ \langle \tilde n^{*} \tilde\delta_i^{t} \rangle 
+ \langle \tilde n_i^{*} \tilde n_i \rangle 
\\
\label{eqn:pii}
&= P_{ii}^{t}(\bar{k}) + P^n_{ii}(\bar{k})
\end{align}
\end{minipage}
\hfill
\begin{minipage}{0.45\textwidth}
\begin{align}
\nonumber
P_{ij}&^m(\bar{k}) = \langle\tilde\delta_{i}^{m *} \tilde\delta_{j}^m\rangle = \langle(\tilde\delta_i^t + \tilde{n}_i)^* (\tilde\delta_j^t + \tilde{n}_j)\rangle \\
\nonumber
&= \langle \tilde\delta_i^{t *} \tilde\delta_j^{t} \rangle 
+ \langle \tilde\delta_i^{t *} \tilde n_j \rangle 
+ \langle \tilde n^{*} \tilde\delta_j^{t} \rangle 
+ \langle \tilde n_i^{*} \tilde n_j \rangle \\
\label{eqn:pij}
&= P_{ij}^{t}(\bar{k})
\end{align}
\end{minipage}

Unlike the auto-power spectrum, the cross-spectrum does not carry a noise power bias in its measurement of the underlying true theoretical signal.

For primarily Gaussian fields, $\langle \delta_1 \delta_2 \delta_3 \delta_4 \rangle = \langle \delta_1 \delta_2 \rangle \langle \delta_3 \delta_4 \rangle + \langle \delta_1 \delta_3 \rangle \langle \delta_2 \delta_4 \rangle + \langle \delta_1 \delta_4 \rangle \langle \delta_2 \delta_3 \rangle$. So, we write the expected variance in the measured power spectrum (given real-valued fields) at a given 3-D $\bar k$:
\begin{align}
\nonumber
\sigma^m_{ij}(\bar{k})^2 &= \langle P_{ij}^{m *} P_{ij}^{m}\rangle^2 - \langle P_{ij}^{m} \rangle^2 =  \big( \langle \tilde\delta_i^{m} \tilde\delta_j^{m *} \tilde\delta_i^{m *} \tilde\delta_j^{m} \rangle - (P_{ij}^m)^2 \big)
\\
\nonumber
\langle \tilde\delta_i^{m} \tilde\delta_j^{m *} \tilde\delta_i^{m *} \tilde\delta_j^{m} \rangle &= 
\langle \tilde\delta_i^{m} \tilde\delta_j^{m *} \rangle \langle \tilde\delta_i^{m *} \tilde\delta_j^{m} \rangle 
+ \langle \tilde\delta_i^{m} \tilde\delta_i^{m *} \rangle \langle \tilde\delta_j^{m *} \tilde\delta_j^{m} \rangle 
+ \langle \tilde\delta_i^{m} \tilde\delta_j^{m} \rangle \langle \tilde\delta_i^{m *} \tilde\delta_j^{m *}  \rangle
\\
\nonumber
&= (P_{ij}^m)^2 + P_{ii}^m P_{jj}^m + (P_{ij}^m)^2
\end{align}

We bin the 3-D power spectrum, and the corresponding measurement uncertainty, within regions of constant $|\bar k|$ (i.e., thin spherical shells of radius $k$). If $N_{ij, \text{modes}}$ is the number of modes (lattice points with $|\bar k| \in (k, k+\delta k)$), then the per-bin variance is an average of  $N_{ij, \text{modes}}$ measurements and carries a factor of ${1}/{2 N_{ij, \text{modes}} (\bar{k})}$. The factor of 2 is included as we assume our density field to be real-valued.

\begin{minipage}{0.3\textwidth}
\begin{align}
\sigma^m_{ii}({k})^2 &= \frac{[P_{ii}^t({k}) + P^n_{ii}({k})]^2}{N_{ij}({k})}
\end{align}
\end{minipage}
\hfill
\begin{minipage}{0.6\textwidth}
\begin{align}
\sigma^m_{ij}({k})^2 &= \frac{1}{2 N_{ij}({k})} \big[ P_{ij}^t({k})^2 + (P_{ii}^t + P^n_{ii})(P_{jj}^t + P^n_{jj}) \big]
\end{align}
\end{minipage}

\subsection{Window Functions \label{appendix:windows}}
Due to finite coverage and resolution, the measured power spectra are not Eqns. \ref{eqn:pii} or \ref{eqn:pij}, but rather the windowed versions:

\begin{minipage}{0.5\textwidth}
\begin{align}
P_{ii}^m(\bar{k}) &= W_{i}(\bar{k}) \times P_{ii}^{t}(\bar{k}) + N_{ii}(\bar{k}),
\end{align}
\end{minipage}
\hfill
\begin{minipage}{0.4\textwidth}
\begin{align}
P_{ij}^m(\bar{k}) &= W_{ij}(\bar{k}) \times P_{ij}^{t}(\bar{k}),
\end{align}
\end{minipage}
with $W_{i, j}(\bar{k}) = |w_{i, j}(\bar{k})|^2$ or $W_{ij}(\bar{k}) = w_{i}(\bar{k})^* w_{j}(\bar{k})$. 

Substituting these expressions instead, the variance in the measurements are (some $\bar{k}$ are suppressed for readability.):

\begin{minipage}{0.25\textwidth}
\begin{align}
\sigma^m_{ii}({k})^2 &= \frac{[W_{i} P_{ii}^t + P^n_{ii}]^2}{N_{ij}}
\end{align}
\end{minipage}
\hfill
\begin{minipage}{0.65\textwidth}
\begin{align}
\sigma^m_{ij}({k})^2 &= \frac{1}{2 N_{ij}} \big[ (W_{ij}P^t_{ij})^2 + (W_{i}P_{ii}^t + P^n_{ii})(W_{j}P_{jj}^t + P^n_{jj}) \big]
\end{align}
\end{minipage}

Equivalently, the variances in the true power spectra are:

\begin{minipage}{0.25\textwidth}
\begin{align}
\sigma^t_{ii}({k})^2 &= \frac{[ P_{ii}^t + P^n_{ii} / W_{i}]^2}{N_{ij}}
\end{align}
\end{minipage}
\hfill
\begin{minipage}{0.65\textwidth}
\begin{align}
\label{eqn:crossnoise}
\sigma^t_{ij}({k})^2 &= \frac{1}{2 N_{ij}} \big[ (P^t_{ij})^2 + (P_{ii}^t + P^n_{ii}/W_{i})(P_{jj}^t + P^n_{jj}/W_{j}) \big]
\end{align}
\end{minipage}

The S/Ns we quote are in the true power spectra:  $\text{S/N}_{ii, jj, ij}(k, z) = \sigma_{ii, jj, ij}^t \big/ P_{ii, jj, ij}^t$.

\bibliographystyle{JHEP}
\bibliography{main}{}

\end{document}